\documentclass[aps,prx,longbibliography,twocolumn,amsmath,amssymb,floatfix,eqsecnum,superscriptaddress]{revtex4-1}

\usepackage{amsmath}
\usepackage{amssymb}
\usepackage{amsthm}
\usepackage[pdftex]{color} 
\usepackage{graphicx}% Include figure files
\usepackage{dcolumn} % Align table columns on decimal point
\usepackage{bm} % bold math
\usepackage{ulem}

\usepackage{arydshln} % package for for introducing dashed lines in matrices

\usepackage[driverfallback=dvipdfm]{hyperref} % So that it can be run with pdftex

\usepackage{longtable}
\usepackage{ulem}   % to strike things out
\newcommand{\be}{\begin{equation}}
\newcommand{\ee}{\end{equation}}

\newcommand{\bsp}{\begin{split}}
\newcommand{\esp}{\end{split}}

\begin{document}

\author{Luiz H. Santos}

\affiliation
{
Department of Physics,  Emory  University, Atlanta,  Georgia  30322,  USA
}

\title{
   Parafermions in Hierarchical Fractional Quantum Hall States
   }

\begin{abstract}
Motivated by recent theoretical progress demonstrating the existence of non-Abelian
parafermion zero modes in domain walls on interfaces between two dimensional Abelian topological phases of matter, 
we investigate the properties of gapped interfaces of 
hierarchical fractional quantum Hall states, in the lowest Landau level,
characterized by the Hall conductance $\sigma_{xy}(m,p) = \frac{p}{2mp+1} \frac{e^{2}}{h}$,
for integer numbers $(m,p)$ with $m,p \geq 1$. 
The case $m=1$ corresponds to the experimentally well established
sequence of fractional quantum
Hall states with
$\sigma_{xy} = \frac{1}{3}\frac{e^{2}}{h},
\frac{2}{5}\frac{e^{2}}{h},
\frac{3}{7}\frac{e^{2}}{h},
...\,
$,
which has been observed in many two dimensional electron gases.
Exploring the mechanism by which the $(m,p+1)$ hierarchical state is generated
from the condensation of quasiparticles of the ``parent" state $(m,p)$, 
we uncover a remarkably rich sequence of parafermions in hierarchical 
interfaces whose quantum dimension $d_{m,p}$ 
depends both upon the total quantum dimension 
$\mathcal{D}_{m,p} = \sqrt{2mp+1}$ 
of the bulk Abelian phase, as well as on the parity of the ``hierarchy level" $p$, 
which we associate with the $\mathbb{Z}_2$ stability of
Majorana zero modes in one dimensional topological superconductors.
We show that these parafermions reside on domain walls separating
segments of the interface where the low energy modes are gapped
by two distinct mechanisms: 
(1) a charge neutral backscattering process
or 
(2)
an interaction that breaks $U(1)$ charge conservation symmetry and 
stabilizes a condensate whose charge depends on $p$. 
Remarkably, this charge condensate corresponds to a clustering of quasiparticles 
of fractional charge $\frac{p}{2mp+1}\,e$, allowing us to draw a 
correspondence between these fractionalized condensates and Read-Rezayi non-Abelian 
fractional quantum Hall cluster states.
\end{abstract}

\maketitle

\date{\today}

%\tableofcontents

%\newpage

\section{Introduction}

Topological phases of matter are promising systems to realize
fault-tolerant quantum computation due to the long-range
entanglement of the quantum many-body state.~\cite{xgwen-book}
Emergent quasiparticles in two-dimensional (2D)
topological phases 
obeying fractional statistics are a potential resource for quantum information science,
particularly so if the system hosts non-Abelian quasiparticles, which allow for the assembling of a degenerate ground state manifold 
where quantum information can be stored and manipulated.~\cite{Nayak-RMP-2008}
Non-Abelian phases have been theoretically investigated in a variety of contexts,
from fractional quantum Hall (FQH) systems~\cite{mooreread91}
to quantum spin liquids~\cite{kitaev-2006}
and recent years have seen exciting experimental
progress to detect signatures of non-Abelian quasiparticles.~\cite{banerjee-2018,kashara-2018}

In the last two decades it has been noticed
that superconductivity is an important mechanism
to stabilize emergent low energy excitations with non-Abelian character. 
A well known example is Kitaev's one-dimensional (1D)
p-wave superconductor supporting Majorana zero modes at the edges.~\cite{Kitaev-2001} 
In this context, the edge of the finite system behaves as a domain wall interpolating between a non-trivial superconductor and a charge neutral insulator, i.e., vacuum.

Recent theoretical breakthroughs in topological phases have
demonstrated that emergent non-Abelian extrinsic defects
can be stabilized as domain wall states in edges and interfaces 
or boundaries of 2D topological phases whose bulk quasiparticles obey solely Abelian statistics.~\cite{BarkeshliQi-2012,Lindner-2012,Clarke-2013,Cheng-2012,Vaezi-2013,BarkeshliJianQi-2013-a,BarkeshliJianQi-2013-b,Mong-2014,khanteohughes-2014,SantosHughes-2017}
In certain cases previously considered, domain walls represent twist defects 
of an anyonic symmetry, which is a transformation that permutes the anyons without changing their fundamental statistical 
properties.~\cite{ostrik-2003,kitaev-2006,BarkeshliWen-2010,Bombin-2010,BeigiShorWhalen-2011,BarkeshliJianQi-2013-a,BarkeshliJianQi-2013-b} 

\begin{figure}
\includegraphics[width=\columnwidth]{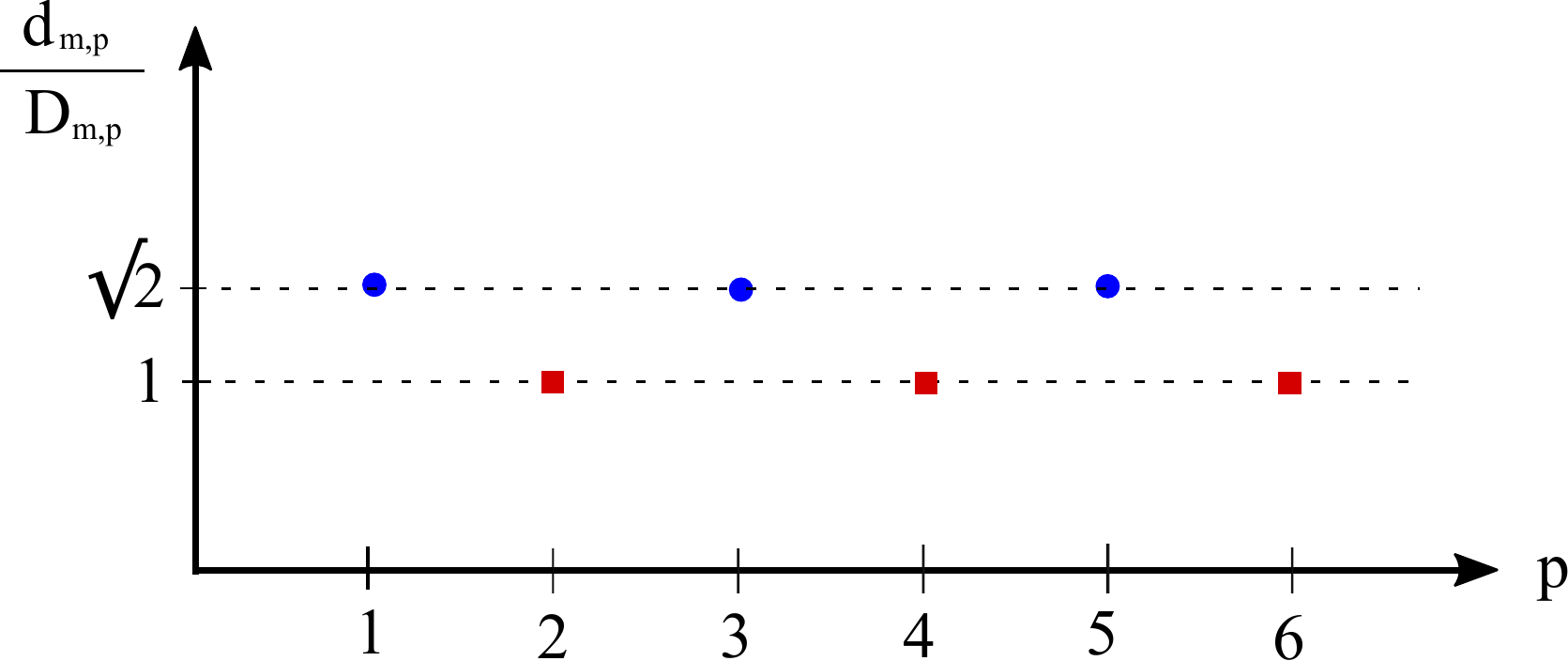}
\caption{
Ratio of the parafermion quantum dimension $d_{m,p}$ to the 
total quantum dimension of the bulk Abelian hierarchical FQH phase $D_{m,p}$
plotted as a function of the hierarchy level $p$,
for a fixed integer $m$.
This plot captures the even-odd dependence
given by Eq.~\ref{eq: d and e relation},
with blue circles and red squares representing,
respectively, odd and even values of $p$.
}
\label{fig:plot-parafermion}
\end{figure}

In the presence of such domain walls, the system can encode a non-trivial ground state degeneracy.
The zero modes in question, which constitute a generalization of Majorana fermions, 
are referred to as parafermions.~\cite{fendley-2012}
Parafermions have been introduced to describe phase transitions of 2D classical clock models 
with $\mathbb{Z}_n$ symmetry~\cite{fradkin-kadanoff-1980,FateevZamolodchikov-1985},
and in recent years, there has been a renewed interest surrounding
the relationship between parafermions and topological systems.~\cite{alicea-fendley-2016} 
A system with $2N$ parafermion zero modes
furnishes a ground state manifold with $d^{2N}$ states,~\cite{comment-parafermion}  
where $d$ represents the quantum dimension of the parafermion.
Majorana fermions correspond to the special 
case $d=\sqrt{2}$, which occur as zero energy excitations
in the edges of 1D topological superconductors~\cite{Kitaev-2001} and
at the vortex core of 2D chiral p-wave superconductors.~\cite{readgreen2000,ivanov-2001}

% \begin{figure}
% \includegraphics[width=\columnwidth]{plot-parafermion-v2.pdf}
% \caption{
% %
% Ratio of the parafermion quantum dimension $d_{m,p}$ to the 
% total quantum dimension of the bulk Abelian hierarchical FQH phase $D_{m,p}$
% plotted as a function of the hierarchy level $p$,
% for a fixed integer $m$.
% %
% This plot captures the even-odd dependence
% given by Eq.~\ref{eq: d and e relation},
% with blue circles and red squares representing,
% respectively, odd and even values of $p$.
% }
% \label{fig:plot-parafermion}
% \end{figure}

A 2D electron gas under external magnetic field
provides a rich realization of Abelian phases of matter
in the form of the FQH effect.~\cite{Prange-book} 
In addition to the Laughlin states~\cite{Laughlin-1983}, 
a plethora of FQH plateaus  
with quantized Hall conductance 
are observed upon changing the magnetic field or the electron density.
A remarkable aspect of these topological phases, particularly
when in the first Landau level, is their hierarchical organization~\cite{Jain1989,Haldane-1983,Halperin-1984}
into a sequence on incompressible states characterized by the 
quantized Hall conductance
\begin{equation}
\label{eq:def hall sequence}
\sigma_{xy}(m,p) = \frac{e^{2}}{h}\,\frac{p}{2mp+1}	
\,,
\end{equation}
where $e$ is the electron's charge, $h$ is the Planck's constant,
$m$ and $p$ are integer numbers greater or equal than one, which characterize
the sequence of FQH states with filling fraction $\nu(m,p) = \frac{p}{2mp+1} < 1$. 
The index $p$ labels the ``position" of the hierarchical state, whose primary
state ($p=1$ and fixed value of $m$) is the
Laughlin state with filling fraction $\nu(m,1) = 1/(2m+1)$.
We shall refer to each of the hierarchical states above by a pair of integer numbers $(m,p)$.
For instance, in the FQH plateaus of Hall conductance $\sigma_{xy}/(e^2/h) = 1/3, 2/5, 3/7, 4/9,...$, 
the condensation of quasiparticles of the primary Laughlin state with $\sigma_{xy}/(e^2/h) = 1/3$
yields the first hierarchical state with $\sigma_{xy}/(e^2/h) = 2/5$;
which in turn gives rise to the second hierarchical state with $\sigma_{xy}/(e^2/h) = 3/7$, and so on.
Even though the investigation of bulk topological properties of the hierarchy of Abelian
FQH in the lowest Landau level has a long history,~\cite{Haldane-1983,Halperin-1984,Blok-Wen-1990,Read-1990,FrohlichZee-1991,WenZee-1992,FrohlichThiran-1994}
the relationship between the bulk anyon condensation 
and the properties of non-Abelian parafermion
zero modes supported by these phases Abelian phases
remains an open problem.
\textit{The goal of this work is to address this problem, 
thus establishing a correspondence between the hierarchy of FQH states
and the local interactions on their interfaces
capable of stabilizing non-Abelian parafermion zero modes.
}
We also note that, by means of a ``folding transformation," 
the results obtained here for the non-chiral interfaces of 
hierarchical time-reversal symmetry breaking FQH states straightforwardly apply 
to the non-chiral edge states pertaining to the hierarchy of 
time-reversal symmetric Abelian Fractional Topological insulators.~\cite{Santos-hierarchy-2011,LevinStern-2012}

In interfaces of Laughlin states with filling fraction $\nu = \frac{1}{2m+1}$ ($m$ integer), 
a domain wall between a segment gapped by the charge neutral backscattering 
$H_{bs} = \psi^{\dagger}_{L}\,\psi_{R} + \textrm{H.c.}$  -- where $\psi_{L}$ and $\psi_{R}$ are, respectively,
the fermionic operators on the left and right edges of the interface -- and 
another region gapped by the charge $2$ condensate $H_{pair} = \psi^{}_{L}\,\psi_{R} + \textrm{H.c.}$,
supports a $\mathbb{Z}_{2(2m+1)}$ parafermion with
quantum dimension $d = \sqrt{2(2m+1)}$.~\cite{Lindner-2012,Clarke-2013,Cheng-2012,Vaezi-2013} 
The $m = 0$ case corresponds to Majorana fermions on domain
walls at the interface of the $\nu=1$ integer quantum Hall state,
akin to the 1D p-wave superconductor.~\cite{Kitaev-2001}
The parafermion zero modes that occur for $m \geq 1$ then represent 
a fractionalization of the Majorana fermion.
A noteworthy aspect of the
quantum dimension of the parafermions 
$d = d_{1D}\times d_{\textrm{bulk}}$, is that it receives a contribution from
the fractionalized bulk Abelian state, $d_{\textrm{bulk}} = \sqrt{2m+1}$, as 
well as a contribution $d_{1D} = \sqrt{2}$ stemming from the $1$D physics.
This property of Laughlin-type interfaces 
raises some important questions regarding 
the remaining sequence of hierarchical states: 
(1) What is the nature of local interactions and domain
walls responsible for parafermions?
(2) Specifically, does a charge 2 pairing interaction
play an important role for generic hierarchical
states, in complete analogy to Laughlin states?
(3) What is the interplay between bulk topology
and 1D physics in determining the quantum
dimension of parafermions?

In this work, we will address these questions
using an effective Luttinger liquid
theory to describe the low energy modes of the
homogeneous interface between hierarchical FQH states.
The presence of chiral edge modes is a well-known
consequence of the topological order of the bulk
hierarchical state described by the bulk Chern-Simons gauge theory.~\cite{Wen-1995}
In this effective edge/interface theory,
local operators that open an energy gap 
are expressed as generalized sine-Gordon (local) operators, 
which satisfy a compatibility condition (or, null condition)
to ensure a stable gapped fixed point.~\cite{Haldane-1995}
Since the number of low energy modes at the interface
grows with the level of the hierarchy $p$, 
any attempt of a general understanding of the parafermion
problem that applies to the entire hierarchical sequence would seem a hopeless question. 
Nevertheless, much to the contrary, 
we shall demonstrate that an general and comprehensive
understanding of the universal properties of 
parafermions is possible
because of the relationship established between 
states of the hierarchy via anyon condensation.
Specifically, the anyon condensation
mechanism translates into a hierarchy of K 
matrices describing the bulk anyons, which, in turn,
will allow us to systematically study the local interactions
that gap the homogeneous interfaces.

By an explicit analysis of the sine-Gordon gap opening 
operators at the homogeneous interface, we shall precisely 
identify local interactions that stabilize parafermion zero modes at domain walls
separating gapped segments where charge conservation
is preserved from other segments
where the interactions break $U(1)$ charge conservation symmetry	
and give rise to a condensate of charge
\begin{equation}
\label{eq:condensate Q - general}
\mathcal{Q}_{m,p} =
\begin{cases}
2p &  p \in \textrm{odd}
\\
p &  p \in \textrm{even}
\end{cases}
\,,
\end{equation}
which, notably, depends on the level $p$ of the hierarchy.
In particular, it shows an even-odd effect as a function of $p$.
Moreover, we observe that, in general, $\mathcal{Q}_{m,p} > 2$, which 
departs from the Laughlin interface.~\cite{Lindner-2012,Clarke-2013,Cheng-2012,Vaezi-2013}
Thus for $p \geq 2$, we identify new forms 
of local interactions that require a 
clustering mechanism beyond the conventional BCS pairing.~\cite{BCS-1957}
Explicit form of these local interactions will be discussed
in Sections~\ref{sec: hierarchy examples}
and~\ref{sec:General parafermion hierarchy}.

Our main result is summarized in Fig.~\ref{fig:plot-parafermion}.
By studying the low energy properties
of the domain walls separating the $U(1)$ symmetry preserving
and broken regions of the interface, we will show that 
they support parafermion zero modes with quantum dimension
\begin{equation}
\label{eq: d and e relation}
d_{m,p}
=
\begin{cases}
\sqrt{2}\,\mathcal{D}_{m,p} &  p \in \textrm{odd}
\\
\mathcal{D}_{m,p} &  p \in \textrm{even}
\end{cases}
\,,
\end{equation}
where $\mathcal{D}_{m,p} = \sqrt{2mp+1}$ is the total quantum dimension of the bulk state,
which is related to the topological entanglement entropy~\cite{kitaev-preskill-2006,levin-wen-2006} of the hierarchical state $(m,p)$
via $\gamma_{m,p} = \log{\mathcal{D}_{m,p}}$ and to the 
minimum quasihole charge $e^{*}_{m,p} = \mathcal{D}^{-2}_{m,p}$.
We again notice the quantum dimension
of the parafermions reflect an even-odd
effect in terms of $p$, similarly to
Eq.~\ref{eq:condensate Q - general}.
Furthermore, the emergence of non-Abelian zero modes
with quantum dimensions given by Eq.~\ref{eq: d and e relation}
suggest a different mechanism than that considered in
Refs.~~\cite{ostrik-2003,kitaev-2006,BeigiShorWhalen-2011,BarkeshliJianQi-2013-a,BarkeshliJianQi-2013-b},
where the defects with non-Abelian character relate to 
twist defects of a symmetry of the anyon group of the Abelian state.
For instance, in Section~\ref{sec: hierarchy examples}
we show that, in the first and second hierarchical states, 
twist defects of the charge conjugation anyonic symmetry behave as a Majorana fermions ($d = \sqrt{2}$)
and twist defects associated with a ``layer permutation" anyonic symmetry~\cite{BarkeshliJianQi-2013-a} 
are trivial ($d = 1$), in contrast with Eq.~\ref{eq: d and e relation}.

Equations~\ref{eq:condensate Q - general} and~\ref{eq: d and e relation}
embody a rich fractionalization phenomenon at the gapped interface.
As shall be demonstrated here,
a segment of the interface where the 
charge condensate Eq.~~\ref{eq:condensate Q - general}
is realized is associated with the expectation
value of an operator of charge $\nu_{m,p} = \frac{p}{2mp+1}$ (recall $e=1$ unit),
which shows that the charge $p$ operator (for $p$ even)
is realized by a cluster of $(2mp+1)$ quasiparticles.
It turns out the appearance of $\mathbb{Z}_{2mp+1}$ parafermions
with quantum dimension $d_{m,p} = \sqrt{2mp+1}$
is a direct consequence of this clustered state,
as shall be explained later. 
In the odd $p$ case, the same type of charge condensate is formed, however,
according to Eq.~\ref{eq: d and e relation}, domain walls support $\mathbb{Z}_{2(2mp+1)} \cong \mathbb{Z}_{2} \oplus \mathbb{Z}_{2mp+1}$ 
parafermions with quantum dimension $d_{m,p} = \sqrt{2}\times\sqrt{2mp+1}$,
where the extra $\mathbb{Z}_2$ structure
is reminiscent of Majorana zero models in 1D topological superconductors.

We argue that this even-odd effect is a manifestation
of the $\mathbb{Z}_2$ classification of 1D topological superconductors,
where the integer index $p$ plays the role of the number of stacked copies of 1D topological superconductors. 
The interpretation of this result is natural in the hydrodynamical Abelian Chern-Simons 
Abelian theory of the hierarchical FQH states, where the universal information of the hierarchical
states $(m,p)$ is represented by a square integer valued 
$K$ matrix of dimension $p$.~\cite{Blok-Wen-1990,Read-1990,FrohlichZee-1991,WenZee-1992,FrohlichThiran-1994} 
Under a suitable SL$(p,\mathbb{Z})$ transformation, the $K$ matrix 
can be interpreted as a $p$-layer FQH system, which
reduces to an integer quantum Hall system of $p$ filled Landau when $m=0$. 
This interpretation of the Chern-Simons hydrodynamical theory
will enable contact with the $\mathbb{Z}_2$ classification of 1D topological superconductors
and support the validity of Eq.~\ref{eq: d and e relation},
which will be explicitly derived in 
Sec.~\ref{sec:General parafermion hierarchy}.

The anyon cluster state realized at
the homogeneous interface of hierarchical FQH states
bears a remarkable resemblance with the Read-Rezayi non-Abelian
states where the ground state is build from
clusters of $k$ electrons which yield a gapped bulk
with non-Abelian quasiparticles and an edge that 
supports a chiral neutral  
$\mathbb{Z}_k$ parafermion mode.~\cite{read-rezayi-1999}
The case $k=2$ corresponds to the Moore-Read paired state~\cite{mooreread91} where electrons (or composite fermions) 
form a paired state whose neutral sector is described by an effective chiral p-wave superconductor.~\cite{readgreen2000}
Reference~\cite{Mong-2014} has shown that 
the $\nu=2/3$ FQH coupled to a superconductor
can support $\mathbb{Z}_3$ parafermion zero modes
on domain walls. Quite remarkably, hybridization of the 
$\mathbb{Z}_3$ parafermions modes throughout the bulk 
can give rise to a non-Abelian phase with 
properties similar to the $\mathbb{Z}_3$ 
Read-Rezayi FQH states that supports non-Abelian
anyons capable of realizing universal quantum computation.~\cite{Mong-2014,stoudenmire-2015}
By the same token, 
and given the generality of the results established here,
we expect that the deconfinement of parafermion zero modes realized in 
the hierarchy of Abelian FQH states to give rise to a rich class
of 2D non-Abelian phases, thus unveiling fresh connections 
between families of Abelian and non-Abelian phases.

This paper is organized as follows. 
In Section~\ref{sec: bulk and edge hierarchy theory},
we give an overview of the 2D hydrodynamical Chern-Simons
theory for the bulk hierarchical Abelian FQH states.~\cite{Blok-Wen-1990,Read-1990,FrohlichZee-1991,WenZee-1992,FrohlichThiran-1994}
The topological information about the hierarchical Abelian state $(m,p)$ 
is encoded by the $K_{m,p}$ matrix and charge vector $q_{m,p}$. 
An fundamental point of this discussion is that
the hierarchy of Abelian states, related by anyon condensation, establishes a useful mapping between K matrices
of the elements of the hierarchy: 
$K_{m,1} \rightarrow K_{m,2} \rightarrow ... \rightarrow K_{m,p-1} \rightarrow K_{m,p} \rightarrow ...$,
with a corresponding mapping for the charge vectors. 
This recursive form of the K matrix and charge vector 
will be explored in order to establish the 
properties expressed in Eq.~\ref{eq: d and e relation} and Eq.~\ref{eq:condensate Q - general}.
In Section~\ref{sec:Gapped interfaces in Hierarchical FQH states}
we provide a general discussion of the properties of domain walls and parafermion
zero modes on the interface of hierarchical
states, where we shall consider two sets of local interactions, one that 
preserves and one that breaks $U(1)$ charge conservation.
An important take home message of this
general discussion is that the quantum dimension of parafermions
can be efficiently
calculated for any hierarchical $(m,p)$ FQH state, 
despite the fact that the number of edge modes scales with the hierarchy level $p$. 
Then, in Section~\ref{sec: hierarchy examples}, we apply this formalism to the Laughlin primary 
states ($p=1$) and the first three hierarchical states $p = 2, 3, 4$. 
(In the case $m=1$, these represent the FQH states with filling
fractions $2/5$, $3/7$ and $4/9$.)
This explicit analysis will be crucial 
in pointing to the general hierarchy of parafermions,
which will be worked out in Section~\ref{sec:General parafermion hierarchy}.
Finally, in Section~\ref{sec:summary_and_discussions}, we shall summarize and discuss our results,
as well as present perspectives for future directions.

\section{Overview of the Hierarchy of Abelian FQH States}
\label{sec: bulk and edge hierarchy theory}

The stability of the sequence of incompressible hierarchical FQH states characterized by 
quantized $\sigma_{xy}(m,p)= \frac{e^2}{h}\frac{p}{2pm+1}$
can be understood in terms of the effective nucleation
of an even number $2m$ of flux quanta per electron -- implemented by a Chern-Simons gauge field~\cite{Lopez1991} --,
giving rise to composite fermions~\cite{Wilczek-1982,Jain1989} which,
at mean field level, occupy an integral number $p$ of filled effective Landau levels.

An alternative description to the composite fermion approach
employs a Chern-Simons hydrodynamical theory
to capture the universal properties of the hierarchical FQH states.~\cite{Blok-Wen-1990,Read-1990,FrohlichZee-1991,WenZee-1992,FrohlichThiran-1994}
This hydrodynamical approach explains the sequence of Abelian FQH states 
in terms of sequential anyon condensations.
Given the $(2+1)$-dimensionality of the problem,
electron and quasiparticle conserved currents
each can be parametrized by a $U(1)$ gauge field.
The condensation of quasiparticles in a given plateau state labeled by $(m,p)$
then gives rise to the hierarchical state $(m,p+1)$, whose effective theory
contains an additional Chern-Simons field.
The resulting effective theory of the $(m,p)$ hierarchical state 
then corresponds to an Abelian Chern-Simons theory that depends
upon $p$ flavors of gauge fields, 
where the Aharonov-Bohm phases associated with exchange of gauge fluxes are encoded
in the integral square and symmetric $K$ matrix of dimension $p$, which characterizes
the Abelian topological order of the hierarchical FQH state $(m,p)$.

In addition to the Abelian statistics of bulk quasiparticles, the hydrodynamical Chern-Simons theory
yields direct information about the low energy properties of the edge states, which form a chiral Luttinger liquid.~\cite{Wen-1995}
The enlargement of the dimension of the $K$ matrix as a function of the hierarchical parameter $p$
signals the increase in the number of chiral edge modes. 
Therefore, when considering local interactions among the modes of such an interface,
it is seen that opening of an energy gap is achieved by generalized sine-Gordon local operators
whose forms are constrained by the $K$ matrix of the bulk state, which in turn
provides a potent link between local operators at the interface and the bulk topological order.

The above-mentioned correspondence between 
the bulk phase and local edge operators will be explored to establish a correspondence between
the hierarchical Abelian states and the parafermions zero modes in their interfaces.   
Given the importance of this formalism, in 
Section~\ref{subsec: Abelian Chern-Simons theory of the bulk hierarchical FQH states}
we review essential elements of the hydrodynamical Chern-Simons theory of 2D hierarchical 
FQH states~\cite{Blok-Wen-1990,Read-1990,FrohlichZee-1991,WenZee-1992,FrohlichThiran-1994} 
leading to the recursive form of the $K$ matrix and charge vector and, later in 
Section~\ref{subsec: Luttinger liquid theory of the hierarchical FQH states},
we make contact the chiral 1D Luttinger liquid theory governing the low energy physics of the edge states.~\cite{Wen-1995}

\subsection{Abelian Chern-Simons theory of the hierarchical FQH states}
\label{subsec: Abelian Chern-Simons theory of the bulk hierarchical FQH states}
Throughout the rest of the paper, we work
in units where $e=\hbar=1$, unless when
when we present formulas for the Hall conductance
where the fundamental constants will be explicitly displayed.

Let us begin with Laughlin states at filling fraction $\nu_{m,1} = 1/(2m+1)$,
which are the primary states $(m,1)$ of the hierarchical sequence. 
Their effective low energy theory is captured by the 2D bulk Chern-Simons Lagrangian
\begin{equation}
L^{2D}_{m,1} = -\frac{2m+1}{4\pi}\,\varepsilon^{\alpha\beta\gamma}a^{1}_{\alpha}\partial_{\beta}a^{1}_{\gamma} + 
\frac{1}{2\pi}\varepsilon^{\alpha\beta\gamma}A_{\alpha}\partial_{\beta}a^{1}_{\gamma}	
\end{equation}
where $a^{1}_{\mu}$ is a dynamical Chern-Simons gauge field,
$A_{\mu}$ is the external electromagnetic field,
Greek indices account for space-time coordinates 
$\{ 0, 1, 2 \} = \{ t, x, y \}$ and the conserved electric current is 
$
J^{\alpha} = \frac{1}{2\pi}\varepsilon^{\alpha\beta\gamma}\partial_{\beta}a^{1}_{\gamma}	
$.
Furthermore, here and throughout, repeated indices are summed over.
Integrating out the Chern-Simons gauge field $a^{1}_{\mu}$ yields the electromagnetic response
\begin{equation}
L^{2D, \textrm{response}}_{m,1} = \frac{1}{4\pi}\frac{1}{2m+1}	\varepsilon^{\alpha\beta\gamma}A_{\alpha}\partial_{\beta}A_{\gamma}
\end{equation}
that encodes the Hall conductance $\sigma_{xy}(m,1) = \frac{e^2}{h}\frac{1}{2m+1}$.

Expressing the quasiparticle's conserved current by
$
j_{2}^{\alpha} = \frac{1}{2\pi}\varepsilon^{\alpha\beta\gamma}\partial_{\beta}a^{2}_{\gamma}	
$, expressed in terms of the gauge field $a^{2}_{\mu}$,
the effective theory of the first hierarchical state $(m,2)$ is given by
\begin{equation}
\label{eq:2nd level CS}
\begin{split}
&\,L^{2D}_{m,2}= -\frac{2m+1}{4\pi}\,\varepsilon^{\alpha\beta\gamma}a^{1}_{\alpha}\partial_{\beta}a^{1}_{\gamma}
+
\frac{1}{2\pi}\varepsilon^{\alpha\beta\gamma}A_{\alpha}\partial_{\beta}a^{1}_{\gamma}
\\
&\,
+\frac{1}{4\pi}\,\varepsilon^{\alpha\beta\gamma}a^{1}_{\alpha}\partial_{\beta}a^{2}_{\gamma}
+\frac{1}{4\pi}\,\varepsilon^{\alpha\beta\gamma}a^{2}_{\alpha}\partial_{\beta}a^{1}_{\gamma}
-\frac{2}{4\pi}\,\varepsilon^{\alpha\beta\gamma}a^{2}_{\alpha}\partial_{\beta}a^{2}_{\gamma}
\,,
\end{split}
\end{equation}
where the first two terms of the second line of Eq.~\ref{eq:2nd level CS}
cone from the minimal coupling $j^{\mu}_{2}a^{1}_{\mu}$ and 
the last term captures the property that, in the mean field state,
the density of quasiparticles $j^{0}$ satisfies 
$
j^{0} = \frac{1}{2}\frac{\nabla\times\boldsymbol{a}^{1}}{2\pi}
$,
implying they condense forming a bosonic Laughlin state with filling $1/2$. 

By introducing the Chern-Simons doublet 
$a^{T}_{\mu} = (a^{1}_{\mu},a^{2}_{\mu})$,
Eq.~\ref{eq:2nd level CS} reads
\begin{equation}
\label{eq:2nd level CS v2}
\begin{split}
&\,
L^{2D}_{m,2} 
=
- \frac{1}{4\pi}\,\varepsilon^{\alpha\beta\gamma}a^{T}_{\alpha}
K_{m,2}
\partial_{\beta}a_{\gamma}
+
q^{T}\frac{1}{2\pi}\varepsilon^{\alpha\beta\gamma}A_{\alpha}\partial_{\beta}a_{\gamma}	
\,,
\\
&\,
K_{m,2}
=
\begin{pmatrix}
2m+1 & -1
\\
-1 & 2
\end{pmatrix}
\,,\quad
q_{m,2} = (1,0)^T
\,.
\end{split}
\end{equation}
Then, integrating out the Chern-Simons fields yields the electromagnetic response
\begin{equation}
L^{2D,\textrm{response}}_{m,2} = \frac{1}{4\pi}\frac{2}{4m+1}	\varepsilon^{\alpha\beta\gamma}A_{\alpha}\partial_{\beta}A_{\gamma}
\,,
\end{equation}
which encodes the Hall conductance $\sigma_{xy}(m,2) = \frac{e^2}{h}\frac{2}{4m+1}$
of the first hierarchical state.

Carrying out these previous steps sequentially generates the 
hydrodynamical Chern-Simons theory of the 
hierarchy of Abelian FQH states~\cite{Blok-Wen-1990,Read-1990,FrohlichZee-1991,WenZee-1992,FrohlichThiran-1994}
\begin{subequations}
\label{eq: hierarchical Chern-Simons - basis 1}
\begin{equation}
\label{eq:p level CS}
\begin{split}
L^{2D}_{m,p+1} 
=
- \frac{1}{4\pi}\,\varepsilon^{\alpha\beta\gamma}a^{T}_{\alpha}
K_{m,p+1}
\partial_{\beta}a_{\gamma}
+
q^{T}\frac{1}{2\pi}\varepsilon^{\alpha\beta\gamma}A_{\alpha}\partial_{\beta}a^{1}_{\gamma}	
\,,
\end{split}
\end{equation}
\begin{equation}
\label{eq: hierarchical K matrix - basis 1}
K_{m,p+1}
=
\begin{pmatrix}
 &  &  &  & 0
\\
 &  & K_{m,p} &  & \vdots
\\
 &  &  &  & 0
\\
 &  &  &  & -1
\\
0 & \ldots & 0 & -1 & 2
\end{pmatrix}
\,,	
\end{equation}
\begin{equation}
\label{eq: hierarchical charge vector - basis 1}
q_{m,p+1} = (q_{m,p}~~ 0)^{T}
\,.	
\end{equation}
The physical mechanism by which the $(m,p+1)$ daughter state
is generated from the condensation of anyons of the $(m,p)$ parent state is 
mathematically manifested in the recursive form of the $K$ matrix Eq.~\ref{eq: hierarchical K matrix - basis 1}
and the charge vector Eq.~\ref{eq: hierarchical charge vector - basis 1}.
Moreover,
\begin{equation}
\textrm{det}(K_{m,p}) = 2mp+1
\,,	
\end{equation}
gives the torus ground state degeneracy of the FQH state
and measures the 
total quantum dimension of the 
Abelian topological phase
\begin{equation}
\mathcal{D}_{m,p} = \sqrt{|\textrm{det}(K_{m,p})|} = \sqrt{2mp+1}
\,.	
\end{equation}
\end{subequations}
Finally, the Hall conductance of the $(m,p)$ state,
obtained from integrating out the Chern-Simons, is given by
\begin{equation}
\sigma_{xy}(m,p) = \frac{e^{2}}{h}\,q^{T}_{m,p}\,K^{-1}_{m,p}\,q_{m,p} = \frac{e^{2}}{h}\, \frac{p}{2mp+1}
\end{equation}

It turns out that, for purpose of studying the 
properties of parafermion zero modes
at the interface of hierarchical states, 
the recursive structure embodied 
in the K matrix and charge vector
in Eq.~\ref{eq: hierarchical Chern-Simons - basis 1} 
will play a central role as shall be discussed in Sections~\ref{sec: hierarchy examples}
and~\ref{sec:General parafermion hierarchy}.

An important consideration is that the topological field theory Eq.~\ref{eq: hierarchical Chern-Simons - basis 1}
is only defined up to an SL($p$, $\mathbb{Z}$) transformation
$
a_{\mu} \rightarrow (W^{T})^{-1}\,a_{\mu}
$,
$
K_{m,p} \rightarrow
W\,K_{m,p}\,W^{T}
$
and
$q_{m,p} \rightarrow W\,q_{m,p}$,
which represents a relabeling of the 
quasiparticles that leaves their statistics unchanged.
This freedom can be explored to 
represent the hierarchical FQH state in the alternative basis~\cite{FrohlichZee-1991}
\begin{subequations}
\label{eq: hierarchical Chern-Simons - basis 2}
\begin{equation}
\label{eq: hierarchical K matrix - basis 2}
\begin{split}
\tilde{K}_{m,p}
&\,=
W_{p}
\,
K_{m,p}
\,
W^{T}_{p}
\\
&\,=
\begin{pmatrix}
2m+1 & 2m & 2m & 2m & \ldots & 2m
\\
2m & 2m+1 & 2m & 2m & \ldots & 2m
\\
2m & 2m & 2m+1 & 2m & \ldots & 2m
\\
\vdots & & & &  
\\
2m & 2m & 2m & \ldots & 2m & 2m+1
\end{pmatrix}	
\,,
\end{split}
\end{equation}
\begin{equation}
\label{eq: hierarchical charge vector - basis 2}
\tilde{q}_{m,p} = W_{p} \, q_{m,p} 
=
(1, 1, \ldots, 1)^{T}	
\,,
\end{equation}
where 
\begin{equation}
W_{p} =
\begin{pmatrix}
1 & 0 & 0 & 0 & \hdots & 0
\\
1 & 1 & 0 & 0 & \ldots & 0
\\
1 & 1 & 1 & 0 & \ldots & 0
\\
\vdots
\\
1 & 1 & 1 & 1 & \ldots & 1 
\end{pmatrix}
\in \textrm{SL}(p,\mathbb{Z})
\,.		
\end{equation}
\end{subequations}
%\red{(reshape matrix)}

The K matrix Eq.~\ref{eq: hierarchical K matrix - basis 2}
has the following appealing interpretation: the diagonal
odd integers $2m+1$ alone represent a system of $p$ layers
of Laughlin $\nu=1/(2m+1)$ states. The even off-diagonal factors of $2m$,
on the other hand, represent bosonic correlations among the $p$ fermionic layers.
Notice the charge vector Eq.~\ref{eq: hierarchical charge vector - basis 2}
denotes that the each layer carries unit charge under the external electromagnetic field.
It will prove useful to explore both representations 
Eq.~\ref{eq: hierarchical Chern-Simons - basis 1} and 
Eq.~\ref{eq: hierarchical Chern-Simons - basis 2} 
when describing the properties of parafermions.

\subsection{Luttinger liquid theory of the hierarchical edge states}
\label{subsec: Luttinger liquid theory of the hierarchical FQH states}

\begin{figure}
\includegraphics[width=\columnwidth]{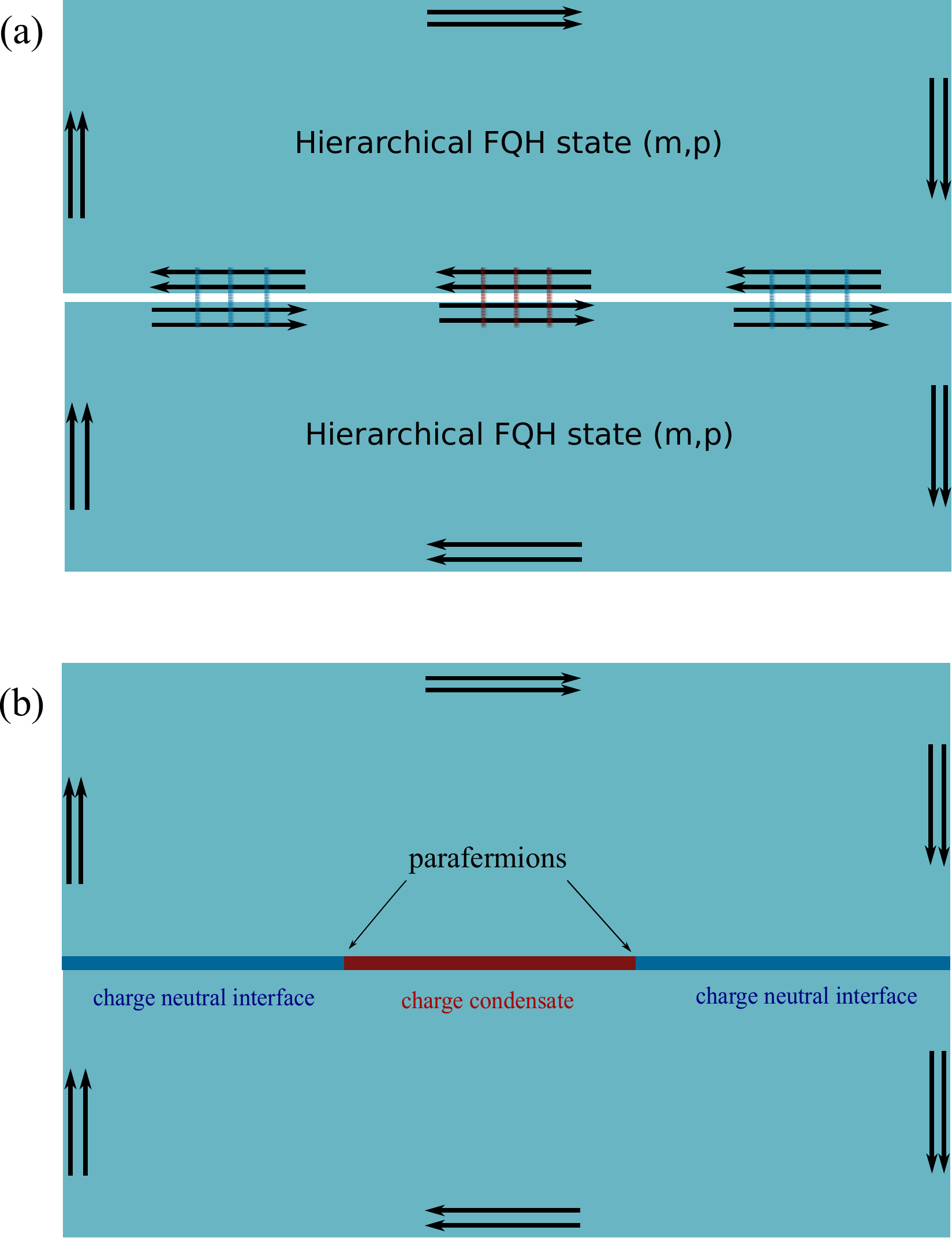}
\caption{(a) Top and bottom panels show the same hierarchical
state, each one supporting a number $p$ of chiral edge modes,
which form a non-chiral interface.
Blue and red dotted lines represent
charge conserving and $U(1)$ symmetry breaking operators
that open an energy gap at the interface.
(b) Blue and red lines at the interface represent segments
gapped by competing mass terms, with domain walls
supporting parafermion zero modes.
}
\label{fig:domains-interface}
\end{figure}

According to the bulk-boundary correspondence,
the bulk topological phase given by Eq.~\ref{eq:p level CS}
supports a chiral edge Luttinger liquid~\cite{Wen-1995}
\begin{equation}
\label{eq:LL hierarchy}
\begin{split}
\mathcal{L}^{R}_{m,p} = &\, -\frac{1}{4\pi} \partial_{t} \Phi^{T}_{R} \cdot K_{m,p} \cdot \partial_{x} \Phi_{R}
-
\frac{1}{4\pi} \partial_{x} \Phi^{T}_{R} \cdot V^{R}_{m,p} \cdot \partial_{x} \Phi_{R}
\\
&\,+
\frac{1}{2\pi}q_{m,p}^{T}\varepsilon^{\alpha\beta}A_{\alpha}\partial_{\beta}\Phi_{R}
\,,
\end{split}
\end{equation}
where $V^{R}_{m,p}$ is a positive-definite matrix ensuring a 
bounded edge spectrum. Furthermore, since all the $p$ eigenvalues of 
$K_{m,p}$ are positive, the edge contains $p$ right-moving modes
described by the fields $\Phi_{R}$,
which are depicted at the bottom part of the interface shown in Fig.~\ref{fig:domains-interface}-(a).
The top part of the interface supports left-moving modes that
are described a similar Lagrangian as in Eq.~\ref{eq:LL hierarchy}, 
albeit with an appropriate sign change of the $K$ matrix that reflects the opposite orientation
of the edge modes.
Therefore, the Luttinger liquid theory of the interface reads
\begin{subequations}
\label{eq: hierarchical interface LL}
\begin{equation}
\label{eq:LL-interface}
\begin{split}
\mathcal{L}_{m,p} = &\, -\frac{1}{4\pi} \partial_{t} \Phi^{T} \cdot \mathcal{K}_{m,p} \cdot \partial_{x} \Phi
-
\frac{1}{4\pi} \partial_{x} \Phi^{T} \cdot \mathcal{V}_{m,p} \cdot \partial_{x} \Phi
\\
&\,+ \frac{1}{2\pi}Q_{m,p}^{T}\varepsilon^{\alpha\beta}A_{\alpha}\partial_{\beta}\Phi
\,,
\end{split}
\end{equation}
where
\begin{equation}
\Phi = 
\begin{pmatrix}
\Phi_{R}
\\
\Phi_{L}
\end{pmatrix}
\,,
\end{equation}
$\Phi_{R}$ and $\Phi_{L}$ are, respectively, the right- and left-moving $p$-tuplet of bosonic edge fields,
\begin{equation}
Q_{m,p}^{T} = 
(q^{T}_{m,p}, ~~ -q^{T}_{m,p})
=
(1, 0, ..., 0, -1, 0,..., 0)	
\end{equation}
is the charge vector and
\begin{equation}
\mathcal{K}_{m,p}
=
\begin{pmatrix}
K_{m,p} & 0
\\
0 & -K_{m,p}
\end{pmatrix}
\,
\end{equation}
\end{subequations}
is the K matrix of the interface. The equal-time commutation relations of the edge fields reads
\begin{equation}
\label{eq: comm relations edge}
[ \partial_{x}\Phi_{a}(x), \Phi_{b}(x') ] = 2\pi i\,\left(\mathcal{K}_{m,p}^{-1}\right)_{ab}\delta(x-x')
\,.
\end{equation}

Gapping the $p$ pairs of counter propagating modes at the interface 
is achieved with a set of $p$ commuting sine-Gordon \textit{local} interactions 
\begin{equation}
\label{eq:generalized interaction LL}
U[\Lambda_{i}] = \cos{(\Lambda^{T}_{i} \mathcal{K}_{m,p}\Phi)} 
\,,
~~ i = 1, ..., p
\,,
\end{equation}
where $\Lambda_{i}$ are $2p$-component integer vectors representing correlated
backscattering processes between right- and left-moving local quasiparticles at the interface.
Furthermore, owing to the non-trivial commutation 
relations, Eq.~\ref{eq: comm relations edge}, satisfied by the edge fields, the integer vectors $\Lambda_{i}$ are required to
satisfy the null condition~\cite{Haldane-1995}
\begin{equation}
\label{eq: null condition interface}
\Lambda^{T}_{i}\,\mathcal{K}_{m,p}\,\Lambda_{j} = 0
\,,~~ i, j = 1, ..., p
\,
\end{equation}
in order for the local interactions Eq.~\ref{eq:generalized interaction LL}
to form a compatible set of mutually
bosonic operators.
Moreover, the integer U$(1)$ charge of the operator 
$U[\Lambda_{i}]$ is
\begin{equation}
\label{eq: charge of operator interface}
\mathcal{Q}[\Lambda_{i}] = \Lambda^{T}_{i} Q_{m,p}
\,.	
\end{equation}

Clearly, integer vectors continue to satisfy Eq.~\ref{eq: null condition interface}
upon rescaling by an integer greater than one.
Then with the respect to the Luttinger liquid
fixed point Eq.~\ref{eq: hierarchical interface LL},
these rescaled null vectors describe local operators with larger 
scaling dimensions, which are then less relevant at low energy
and can be disregarded.
Therefore, in the remaining of this paper,
we shall only focus our attention on null vectors that 
are primitive.~\cite{LevinStern-2012}
As shown in that work, a single integer vector is primitive when the greatest common divisor of its entries is $1$.
A set of integer vectors is primitive if and only if the greatest common divisor of the set of 
minors of the $p \times 2p$ integer matrix $\mathcal{M}[\{\Lambda\}]$ is 1,
where $\mathcal{M}[\{\Lambda\}]$ is the integer matrix whose rows are formed by the null vectors 
$\Lambda_{i}$.

In general, the low energy modes of the interface can 
become gapped due to distinct types of local 
interactions, each one associated with a primitive null set $\{ \Lambda_{i} \}$
satisfying Eq.~\ref{eq: null condition interface}. 
As the number $p$ of counter-propagating modes at the interface grows 
(i.e., as one moves ``deeper" into the hierarchical sequence), 
one expects a corresponding increase 
in the number of gapping channels of the interface, 
as a consequence of more available 
backscattering channels amongst the 
counter-propagating modes.
As such, the investigation of the low energy properties
of interfaces of hierarchical FQH states poses a very 
rich physics problem. In the following discussion,
we shall concentrate on 
certain classes of local interactions
leading to gapped interfaces, 
whose properties will be described in generality
in Section~\ref{sec:Gapped interfaces in Hierarchical FQH states}
and, more specifically, in Section~\ref{sec: hierarchy examples}
and~\ref{sec:General parafermion hierarchy}.

\section{Domain walls in Hierarchical interfaces: General Properties}
\label{sec:Gapped interfaces in Hierarchical FQH states}

We now discuss the properties of domain walls and parafermion zero modes
associated with hierarchical interfaces.
One of the central points of this Section is the ansatz Eq.~\ref{eq: integer vectors and interactions SC}
that describes the $U(1)$ symmetry broken interactions at the interface and which will permit 
us to determine, efficiently, the quantum dimension of the parafermions localized on the domain walls
for an interface that holds $p$ counter propagating modes.
As described in Section~\ref{subsec: Luttinger liquid theory of the hierarchical FQH states}, 
the number of chiral edge modes grows with the hierarchy index $p$,
which increases the number of gap opening channels.
In this context, addressing all possible forms of gapped interfaces seems a formidable task,
which is beyond the scope of this work. Instead, we shall 
focus on a specific class of local interactions, which will be 
shown to stabilize parafermion zero modes on domain walls along the interface.
We shall consider two types of gapped interfaces. The first one is formed
by charge neutral backscattering, while the second one breaks charge conservation. 
Our focus is then on the low energy properties of domain walls separating 
charge conserving and the non-conserving gapped segments.
We note that superconducting pairing correlations have been 
recently induced in integer quantum Hall edges,~\cite{Amet-2016,Lee-2017}
which represents a promising step to create superconductor/FQH heterostructures.
The homogeneous interface described by Eq.~\ref{eq: hierarchical interface LL}
admits
local charge neutral backscattering that gap the interface
and heal the bulk states, as represented 
by the the red segments in Fig.~\ref{fig:domains-interface}-(b).
This interface, which allows Abelian anyons to hop across 
and propagate as \textit{bona fide} deconfined bulk quasiparticles,
is created by the backscattering terms
\begin{subequations}
\label{eq: interactions and null vectors - charge conserving interface}
\begin{equation}
\label{eq: interactions charge conserving interface}
U[\Lambda^{(0)}_{i}] = \cos{(\Lambda^{(0)}_{i} \mathcal{K}_{m,p}\Phi)}
\,,
\end{equation}
where
\begin{equation}
\label{eq:charge conserving null vectors}
\Lambda^{(0)}_{i} = 
\begin{pmatrix}
e_i
\\
e_i
\end{pmatrix}
\,,~~
e_{i} = (0,\ldots,0,\underbrace{1}_{i},0,\ldots,0)^{T}	
%\,,\quad i = 1, ..., p
\end{equation}
\end{subequations}
for $i = 1, ..., p$ is a set of integer vectors. 
Charge conservation obeyed by the interactions Eq.~\ref{eq: interactions charge conserving interface} follows from 
\begin{equation}
\label{eq: charge conservation}	
\begin{split}
\mathcal{Q}[\Lambda^{(0)}_{i}] 
&\,= 
\Lambda^{(0)}_{i} Q_{m,p} 
\\
&\,=
\begin{pmatrix}
e^{T}_{i} & e^{T}_{i}
\end{pmatrix}
\begin{pmatrix}
q_{m,p}
\\
-q_{m,p}
\end{pmatrix}
=
0
\,, 
\end{split}
\end{equation}
and the null condition of the integer vectors Eq.~\ref{eq:charge conserving null vectors} 
\begin{equation}
\label{eq:null condition charge conserving vecs}
\begin{split}
\Lambda^{(0)}_{i}\,\mathcal{K}_{m,p}\,\Lambda^{(0)}_{j}
&\,=
\begin{pmatrix}
e^{T}_i & e^{T}_i
\end{pmatrix}
[K_{m,p} \oplus (-K_{m,p})]
\begin{pmatrix}
e_j
\\
e_j
\end{pmatrix}
\\
&\,=
e^{T}_i\,K_{m,p}\, e_{j}
-
e^{T}_i\,K_{m,p}\,e_{j}
=0
\,
\end{split}
\end{equation}
is verified
$\forall~ i, j = 1, \ldots, p$.
\begin{subequations}

We now consider another set of local interactions
\begin{equation}
\label{eq: interactions SC}
U[\Lambda_{i}] = \cos{(\Lambda_{i} \mathcal{K}_{m,p}\Phi)}
\,
\end{equation}
that break charge conservation 
and depend upon the integer vectors
\label{eq: integer vectors and interactions SC}
\begin{equation}
\label{eq: integer vectors SC}
\begin{split}
\Lambda_{i}
=
\begin{cases}
\Lambda_1 \neq \Lambda^{(0)}_{1} & \textrm{and}~~ \Lambda_1 Q_{m,p} \neq 0
\\
\Lambda^{(0)}_{i} & i = 2, ..., p
\, 
\end{cases}		
	\end{split}		
\end{equation}
\end{subequations}
for $i = 1, \ldots, p$.

The interaction $U[\Lambda_{1}]$ breaks charge conservation, 
while the $U[\Lambda_{i \neq 1}]$ conserve charge.
Despite its simple form, this ansatz will be shown to embody 
a non-trivial charge condensate that gaps the interface 
and stabilizes parafermions on domain walls between segments of the interface 
gapped by the interactions Eq.~\ref{eq: interactions SC}
from those segments gapped by the charge neutral interactions Eq.~\ref{eq: interactions charge conserving interface}.
Furthermore it permits an analytical understanding
of the mechanism behind the formation of domain wall parafermions. 
Since the subset of $p-1$ null vectors $\{ \Lambda_{i} = \Lambda^{(0)}_{i}\,,~~ i = 2 , ..., p \}$
in Eq.~\ref{eq: integer vectors SC} satisfies the null condition, 
Eq.~\ref{eq: null condition interface}
reduces to $p$ independent equations that 
can be solved exactly, as we shall demonstrate in the 
following Sections~\ref{sec: hierarchy examples}
and~\ref{sec:General parafermion hierarchy}.

In order to determine the quantum dimension of the 
parafermion zero modes, we consider a series of domain walls at the interface that separate segments 
$
S_{0} = \cup_{i}\,(  x_{2i}+\varepsilon, x_{2i+1}-\varepsilon)
$
gapped by the interactions Eq.~\eqref{eq: interactions charge conserving interface} 
from the segments 
$
S = \cup_{i}\,(  x_{2i-1}+\varepsilon, x_{2i}-\varepsilon)
$
gapped by the interactions Eq.~\eqref{eq: interactions SC},
where $\varepsilon = 0^{+}$ is a positive regulator for the domain walls.
In the strong coupling limit,
the ground state is obtained by
locking the
sine-Gordon 
terms Eq.~\ref{eq: interactions charge conserving interface}
and Eq.~\ref{eq: interactions SC}
to their minima on the respective segments $S^{(0)}$ and $S$.
The ground state degeneracy can be obtained 
by constructing a set of operators with support on these gapped segments 
\begin{subequations}
\label{eq:gammas and N}
\begin{equation}
\label{eq: Gamma1}
\Gamma_{2i-1,2i} = 
\exp{\Big( \frac{i}{N_{m,p}}\,\int^{x_{2i}+\varepsilon}_{x_{2i-1}-\varepsilon}\,dx\,\Lambda^{(0)}_{1}\mathcal{K}_{m,p}\partial_{x}\Phi  \Big)}
\,,
\end{equation}
\begin{equation}
\label{eq: Gamma2}
\Gamma^{}_{2i,2i+1} = \exp{\Big( \frac{i}{N_{m,p}}\,\int^{x_{2i+1}+\varepsilon}_{x_{2i}-\varepsilon}\,dx\,
\Lambda^{}_{1}\mathcal{K}_{m,p}\partial_{x}\Phi  \Big)}
\,,
\end{equation}
where 
\begin{equation}
\label{eq: d2 general}
N_{m,p} = \Lambda^{(0)}_{1}\,\mathcal{K}_{m,p}\,\Lambda_{1} \in \mathbb{Z}^{*}
\,.
\end{equation}
\end{subequations}
It follows from the commutation relations Eq.~\ref{eq: comm relations edge} that operators
defined in Eq.~\ref{eq: Gamma1} and Eq.~\ref{eq: Gamma2} 
commute with the Hamiltonian along the interface and satisfy the algebra
\begin{equation}
\label{eq: commutation relations non-local ops}
\begin{split}
\Gamma_{2i-1,2i}\,\Gamma_{2j,2j+1}
&\,=
e^{i\,\frac{2\pi}{N_{m,p}}\left( \delta_{i,j} - \delta_{i-1,j}  \right)}
\,
\Gamma_{2j,2j+1}\,\Gamma_{2i-1,2i}	
\\
\Gamma^{N_{m,p}}_{2k-1,2k} &\,= \Gamma^{N_{m,p}}_{2k,2k+1} = 1
\,.
\end{split}
\end{equation}
The dimension of the minimum representation
of this algebra clearly corresponds to the ground state degeneracy.
$\Gamma_{2i,2i+1}$ act as raising or lowering operator to its
neighbors $\Gamma_{2i-1,2i}$ and $\Gamma_{2i+1,2i+2}$, 
as $\mathbb{Z}_{N_{m,p}}$ clock operators. 
For a configuration with $2\,n_{dw}$ domain walls, 
Eq.~\eqref{eq: commutation relations non-local ops}
conveys the ground state degeneracy 
$|N_{m,p}|^{n_{dw}}$
and the quantum dimension of the parafermion
\begin{equation}
d_{m,p} = \sqrt{|N_{m,p}|}	
\,.
\end{equation}
Therefore, the quantum dimension of the parafermions 
depends upon a single integer given by Eq.~\ref{eq: d2 general}.

Finally, the ground state degeneracy
stems from the existence of parafermion zero modes on the domain walls
\begin{equation}
\label{eq:parafermions}
\begin{split}
&\,
\alpha_{2i} = 
e^
{
\frac{i}{N_{m,p}}
\left[\Lambda_1 \, \mathcal{K}_{m,p} \, \Phi(x_{2i}-\varepsilon)
+
\Lambda^{(0)}_1 \, \mathcal{K}_{m,p} \, \Phi(x_{2i}+\varepsilon)
\right]
}
\\
&\,
\alpha_{2i+1} = 
e^
{
\frac{i}{N_{m,p}}
\left[\Lambda^{(0)}_1 \, \mathcal{K}_{m,p} \, \Phi(x_{2i+1}-\varepsilon)
+
\Lambda_1 \, \mathcal{K}_{m,p} \, \Phi(x_{2i+1}+\varepsilon)
\right]
}
\end{split}	
\,,
\end{equation}
which satisfy the $\mathbb{Z}_{N_{m,p}}$ parafermion algebra
\begin{equation}
\alpha_{i}\,\alpha_{j} = e^{i \frac{2\pi}{N_{m,p}}\textrm{sgn}(i-j)}\alpha_{j}\,\alpha_{i}	
\,,
\end{equation}
and are related to the operators in Eq.~\ref{eq:gammas and N} by
\begin{equation}
\alpha^{\dagger}_{2i}\,\alpha_{2i+1}\, \sim \Gamma_{2i+1,2i}
\,,
\end{equation}
with similar relations holding for the other segments on the interface.
This establishes that the bilinear terms constructed out of the 
parafermion operators commute with the Hamiltonian at the interface.

In the following Section we shall impose the null and primitive conditions
to the interactions given by the ansatz Eq.~\ref{eq: integer vectors and interactions SC}
and use it to obtain the local $U(1)$ symmetry broken interactions and parafermion zero modes
for the first few hierarchical states.
(For completeness we shall also revisit the primary Laughlin state
studied in Refs.~\cite{Lindner-2012,Clarke-2013,Cheng-2012,Vaezi-2013}.)
The analysis of these explicit cases will point 
to the general formulation valid for all hierarchical states,
which we shall present in Section~\ref{sec:General parafermion hierarchy}.

\section{Hierarchy of Parafermions: Examples} % (fold)
\label{sec: hierarchy examples}

In this Section we apply the formalism
introduced in Section~\ref{sec:Gapped interfaces in Hierarchical FQH states}
from the primary Laughlin states up until the third hierarchical state.

\subsection{Primary Laughlin state $(m,1)$}
\label{sec: primary states}

The Laughlin state with filling fraction $\nu_{m,1} = \frac{1}{2m+1} = \frac{1}{3}, \frac{1}{5}, ...$ 
is described by a one-component quantum fluid with $K_{m,1} = 2m+1$ and $q_{m,1}=1$. 
The interface Luttinger liquid theory 
Eq.~\ref{eq: hierarchical interface LL} has
\begin{equation}
\label{eq: interface K and Q primary}
\mathcal{K}_{m,1} = 
\begin{pmatrix}
2m+1 & 0
\\
0 & -(2m+1)
\end{pmatrix}
\,,
\quad
Q_{m,1} = (1,-1)^{T} 
\,,
\end{equation}
with electron operators given by $\psi_{L/R} = e^{i(2m+1)\phi_{L/R}}$,
where $\phi_{L/R}$ are the chiral boson fields at the interface.

The charge neutral backscattering Eq.~\ref{eq: interactions and null vectors - charge conserving interface}
for the Laughlin interface reads
\begin{equation}
\label{eq:null-charge-primary}
\begin{split}
&\,
\Lambda^{(0)}_{1} = (1,1)
\,,
\end{split}
\end{equation}
\begin{equation}
\label{eq:interaction charge-primary}
\begin{split}
U[\Lambda^{(0)}_{1}] &\,= 
\frac{\lambda_1}{2}(\psi^{\dagger}_{L}\psi_{R} + \textrm{H.c.}) 
\\
&\,= 
\lambda_{1}\cos{\left[(2m+1)(\phi_{R}-\phi_{L}) \right]}
\,.
\end{split}
\end{equation}

Alternatively, the interface can be gapped via 
charge $2$ electron pairing.
\begin{equation}
\label{eq:null-sc-primary}
\begin{split}
&\,
\Lambda^{}_{1} = (1,-1)
\,,
\end{split}
\end{equation}
\begin{equation}
\label{eq:interaction sc-primary}
\begin{split}
U[\Lambda^{}_{1}] &\,= 
\frac{\lambda^{'}_1}{2}(\psi_{L}\psi_{R} + \textrm{H.c.}) 
\\
&\,= 
\lambda^{'}_{1}\cos{\left[(2m+1)(\phi_{R}+\phi_{L}) \right]}
\,.
\end{split}
\end{equation}

Then, according to Eq.~\ref{eq: d2 general}, 
\begin{equation}
N_{m,1} = \Lambda^{(0)}_{1}\mathcal{K}_{m,1}\Lambda_{1} = 2(2m+1)
\end{equation}
establishes $\mathbb{Z}_{2(2m+1)} \cong \mathbb{Z}_{2} \oplus \mathbb{Z}_{2m+1}$ parafermions 
\begin{equation}
\alpha_{i}\,\alpha_{j}
=
e^{i \frac{2\pi}{2(2m+1)}\textrm{sgn}(i-j)}
\alpha_{j}\,\alpha_{i}
\,
\end{equation}
with quantum dimension
\begin{equation}
d_{m,1} = \sqrt{2}\times\sqrt{2m+1}	 = \sqrt{2}\times \mathcal{D}_{m,1}
\,,
\end{equation}
which are localized on the domain
walls between regions gapped by the charge neutral local backscattering
Eq.~\ref{eq:interaction charge-primary} and regions gapped by the charge $2$ paring
Eq.~\ref{eq:interaction sc-primary}.~\cite{Lindner-2012,Clarke-2013,Cheng-2012,Vaezi-2013}.
Notice that the quantum dimension of the parafermion
shows a contribution from the bulk topological order through 
the total quantum dimension $\mathcal{D}_{m,1} = \sqrt{2m+1}$
of the Laughlin state, as well as a contribution $\sqrt{2}$
reminiscent of Majorana zero modes in a 1D
topological superconductor.~\cite{Kitaev-2001}. 
Because of this 1D effect, even in the absence of deconfined bulk anyons, 
which corresponds to the $\nu=1$ IQH state where $m=0$, there is one Majorana zero mode 
localized on each domain wall of the IQH interface.

The presence of $\mathbb{Z}_{2(2m+1)}$ parafermions
on the domain walls, as seen by 
Eq.~\ref{eq:parafermions}, 
manifests that the charge $1/m$ operator 
\begin{equation}
\langle\, \mathcal{O}_{\frac{1}{2m+1}} \,\rangle
=
\langle\, e^{i\,\frac{(2m+1)(\phi_{R}+\phi_{L})}{2(2m+1)}}\,\rangle
=
\langle\, e^{i\,\frac{(\phi_{R}+\phi_{L})}{2}}\,\rangle
\neq 0
\end{equation}
acquires a non-zero expectation on the segments of the interface that are
gapped by the interaction Eq.\ref{eq:interaction sc-primary}. 

\subsection{First Hierarchical State $(m,2)$}
\label{sec: level 1}

The interface of the first hierarchical state with filling fraction
$\nu_{m,2} = \frac{2}{4m+1} = \frac{2}{5}, \frac{2}{9}, ...$ 
contains two pairs of counter-propagating fields $\Phi^{T} = (\phi^{R}_{1},\phi^{R}_{2},\phi^{L}_{1},\phi^{L}_{2})$.
The Luttinger liquid Lagrangian of the interface, Eq.~\ref{eq: hierarchical interface LL}, has
\begin{equation}
\label{eq: interface K and Q level 1}
\begin{split}
&\,
\mathcal{K}_{m,2} = K_{m,2} \oplus (-K_{m,2})
\,,
\quad
Q_{m,2}^{T} = (q^{T}_{m,2},-q^{T}_{m,2}) 
\\
&\,
K_{m,2} = \begin{pmatrix}
2m+1 & -1
\\
-1 & 2
\end{pmatrix}
\,,\quad
q_{m,2}^{T} = (1,0)
\,.
%\,,
\end{split}
\end{equation} 
where we adopt the hierarchical representation Eq.~\ref{eq: hierarchical K matrix - basis 1}
for the $K$ matrix of the bulk state.

According to Eq.~\ref{eq: interactions and null vectors - charge conserving interface}, 
the pair of local interactions
\begin{subequations}
\label{eq:interactions-charge-level 1}
\begin{equation}
\label{eq:interaction1 charge-level 1}
\begin{split}
U[\Lambda^{(0)}_{1}] &\,= 
\frac{\lambda_1}{2}(\psi^{\dagger}_{1L}\psi_{1R} + \textrm{H.c.}) 
\\
&\,= 
\lambda_{1}\cos{\left[(2m+1)(\phi_{1R}-\phi_{1L}) - (\phi_{2R} - \phi_{2L}) \right]}
\end{split}
\end{equation}
and
\begin{equation}
\label{eq:interaction2 charge-level 1}
\begin{split}
U[\Lambda^{(0)}_{2}]
&\,= 
\frac{\lambda_2}{2}(\psi^{\dagger}_{2L}\psi_{2R} + \textrm{H.c.})
\\
&\,=
\lambda_{2}\cos{\left[-(\phi_{1R}-\phi_{1L}) + 2(\phi_{2R} - \phi_{2L}) \right]}
\,,
\end{split}
\end{equation}
associated with the null vectors
\begin{equation}
\label{eq:null-charge-level 1}
\begin{split}
&\,
\Lambda^{(0)}_{1} = (1,0,1,0)
\,,\quad
\Lambda^{(0)}_{2} = (0,1,0,1)
\,,
\end{split}
\end{equation}
\end{subequations}
gap the interface without breaking charge conservation.
The local operators at the interface 
correspond to $\psi_{a,R/L} = e^{i\,\sum_{b}(K_{R/L})_{ab}\phi_{b,R/L}}$, for $a=1,2$.
Eq.~\ref{eq:interactions-charge-level 1} represents local charge neutral backscattering
that localizes the interface low energy modes.

We now seek the null vectors and corresponding
charge non-conserving interactions that gap the modes of the interface. Following 
Eq.~\ref{eq: integer vectors and interactions SC}, we consider
\begin{equation}
\label{eq:null-sc-level 1}
\begin{split}
&\,
\Lambda_{1} = (x_1,y_1,x_2,y_2)
\,,\quad
\Lambda_{2} = (0,1,0,1)
\,,
\end{split}
\end{equation}
where $x_{1}, x_{2}, y_{1}, y_{2}$ are integers. Notice that the interaction 
$U[\Lambda_{2}] = \cos{(\Lambda_{2}\mathcal{K}_{m,2}\Phi)}$ is charge neutral,
while $U[\Lambda_1]$ is an operator of charge 
$\mathcal{Q}[\Lambda_{1}] = \Lambda_{1}Q_{m,2} = x_1 - x_2$,
and we consider $x_1 \neq x_2$ in what follows.

The null condition satisfied by the integer vectors Eq.~\ref{eq:null-sc-level 1} reads
\begin{subequations}
\begin{equation}
\label{eq:null 1 p=2}
\begin{split}
&\,
\Lambda_{1} \mathcal{K}_{m,2} \Lambda_{1} =
\\
&\, 
(2m+1)( x^2_{1} - x^2_{2}) + 2 \left(y^2_{1}- y^2_{2} - x_1 y_1 + x_2 y_2 \right)
=0
\,,
%%%
\end{split}	
\end{equation}
\begin{equation}
\label{eq:null 2 p=2}
\Lambda_{1}	\mathcal{K}_{m,2} \Lambda_{2} = -x_1+x_2 + 2 y_1 - 2 y_2 = 0
\,.
\end{equation}
\end{subequations}
Solving for $x_1$ in Eq.~\ref{eq:null 2 p=2} and substituting
into Eq.~\ref{eq:null 1 p=2} gives 
\begin{equation}
\label{eq:null 1 p=2 simplified}
(y_1-y_2) (x_2+y_1-y_2) = 0
\,.
\end{equation}

Whereas $y_1 = y_2=y$ and $x_1 = x_2=x$ solves the null conditions, 
it corresponds to a charge neutral null vector 
$\Lambda_{1} = x\Lambda^{(0)}_{1} + y\Lambda^{(0)}_{2}$
that is a linear combination of those in Eq.~\ref{eq:null-charge-level 1}, in which case
the integer vectors Eq.~\ref{eq:null-charge-level 1} and Eq.~\ref{eq:null-sc-level 1} represent the 
same type of gapped interface.
A non-trivial solution, however, corresponds to 
$y_1 - y_2 = t$, 
and
$x_1 = -x_2 = t$
for $t \neq 0$,
such that 
$\Lambda_{1} = (t, t+y_2,-t,y_2)$ for $y_2 \in \mathbb{Z}$.
Moreover, the minors of $\mathcal{M}[\{ \Lambda_1, \Lambda_2 \}]$,
are given by
$
\{t,0,t,t,t,-t\}
$,
such that the primitive condition requires $t = \pm 1$. 
Finally, setting $t=1$ and $y_2 = 0$ yields
\begin{equation}
\Lambda_{1} = (1, 1, -1,0)
\,,	
\end{equation}
and the gap opening interaction
\begin{equation}
\label{eq: interaction sc level 1}
\begin{split}
U[\Lambda_{1}]
&\,=
\frac{u}{2}\,(\psi_{1R}\psi_{2R}\psi_{1L} + \textrm{H.c.})
\\
&\,
=
u\cos{(\Lambda_{1} \mathcal{K}_{m,2}\Phi)}
\\
&\,
=
u\cos{(2m\phi^{R}_{1}  + \phi^{R}_{2} + (1+2m)\phi^{L}_{1} -\phi^{L}_{2})}
\,.		
	\end{split}	
\end{equation}
This interaction 
%Eq.~\ref{eq: interaction sc level 1} 
represents a charge $2$ condensate,
where
$\psi_{1R}$ and $\psi_{1L}$ are both charge $1$ fermionic operators, 
and
$\psi_{2R}$ accounts for a charge zero operator with bosonic self-statistics.

According to Eq.~\ref{eq: d2 general},
\begin{equation}
N_{m,2} = \Lambda^{(0)}_{1}\mathcal{K}_{m,2}\Lambda_{1} = 4m+1
\,
\end{equation}
shows the presence of $\mathbb{Z}_{4m+1}$ parafermions
\begin{equation}
\alpha_{i}\,\alpha_{j}
=
e^{i \frac{2\pi}{4m+1}\textrm{sgn}(i-j)}
\alpha_{j}\,\alpha_{i}
\,
\end{equation}
with quantum dimension
\begin{equation}
d_{m,2} = \sqrt{4m+1} = \mathcal{D}_{m,2}
\,,
\end{equation}
which are localized on the domains of the interface.
This result shows that 
the quantum dimension of the parafermion
is a direct manifestation of bulk topological order 
of the first hierarchical state $(m,2)$
through its total quantum dimension $\mathcal{D}_{m,2} = \sqrt{4m+1}$.

As an application of our analysis, 
the homogeneous interface of
$\nu=2/5$ FQH states (where $m=1$ and $p=2$) supports $\mathbb{Z}_{5}$ parafermions. 
In this case, parafermion operators are constructed from operators
that create fractional charge $2/5$ and charge zero quasiparticle pairs 
on each side to the domain walls. Interestingly, the charge $2$ condensate 
results from the coalescence of a quintuplet of quasiparticles of charge $2/5$. 

The generalization of the $\nu = 2/5$ state to other first hierarchical states
with arbitrary values of $m$ shows that the $\mathbb{Z}_{4m+1}$ parafermion
results from the the formation of a $(4m+1)$-tuplet of charge $2/(4m+1)$ fractional quasiparticles
giving rise to a charge $2$ condensate. This, in turn, is a manifestation 
of the non-zero expectation value of the charge $\frac{2}{4m+1}$ operator
\begin{equation}
\Big\langle\,
\mathcal{O}_{\frac{2}{4m+1}}
\,\Big\rangle 
\equiv
\Big\langle\,
e^{i \frac{\Lambda_{1}\,\mathcal{K}_{m,2}\,\Phi}{4m+1}}
\,\Big\rangle 
\neq 0
\,	
\end{equation} 
on the segments of gapped by the interaction Eq.~\ref{eq: interaction sc level 1}.

Compared with the primary Laughlin states discussed in Section~\ref{sec: primary states},
the quantum dimension of the parafermions in the first hierarchical state
does not manifest a $\sqrt{2}$ contribution expected for the 1D topological superconductor.
To shed light on this result, 
we make use of the representation 
of the first hierarchical state 
Eq.~\ref{eq: hierarchical Chern-Simons - basis 2}
\begin{equation}
\label{eq: K matrix first hierarchical basis 2}
\begin{split}
&\,
\tilde{K}_{m,2}
=
W_{2}\,K_{m,2}\,W^{T}_{2}
=
\begin{pmatrix}
2m+1 & 2m
\\
2m & 2m+1
\end{pmatrix}
\\
&\,
\tilde{q}_{m,2} = (1,1)
\,,
\end{split}
\end{equation}
where the first hierarchical 
state can be thought of as a coupled FQH bilayer,
with each layer carrying U$(1)$ charge $q = 1$. 
Therefore, the degrees of freedom at the interface constitute \textit{two}
pairs of counter-propagating fermion modes, -- twice the number of degrees of freedom in the interface of the primary Laughlin states.
Our analysis then shows
that adding an extra pair of counter-propagating 
modes renders the 
Majorana zero mode unstable, which is an
indication of the $\mathbb{Z}_2$ stability of Majorana fermions
in 1D topological superconductors.

We now draw an important comparison between the
parafermion zero modes discussed in our set up
and the theory of extrinsic defects associated
with anyonic symmetries of the Abelian phase.
In Ref.~\cite{BarkeshliJianQi-2013-a}, the $\mathbb{Z}_2$
twisted defects associated with the layer permutation 
of the Abelian phase characterized by the K
matrix
\begin{equation}
\label{eq:K matrix bilayer m l}
K_{m,\ell}
=
\begin{pmatrix}
m & \ell
\\
\ell & m
\end{pmatrix}
\,	
\end{equation}
where studied and it was shown that domain walls separating 
two distinct gapping charge conserving gap terms support parafermions
with quantum dimension $d_{\mathbb{Z}_2} = \sqrt{|m-\ell|}$.
Comparing Eq.~\ref{eq: K matrix first hierarchical basis 2}
and
Eq.~\ref{eq:K matrix bilayer m l}
shows that that $\mathbb{Z}_2$ twist
defect of the first hierarchical state
is trivial, since $d_{\mathbb{Z}_2}=1$. 
Furthermore, it can be demonstrated~\cite{comment-chargeconjugation}
that the twist defects associated with charge conjugation
anyonic symmetry correspond Majorana fermions,
but not the $\mathbb{Z}_{4m+1}$ parafermions discussed here.

These preliminary findings regarding the 
primary and first hierarchical states
point to the existence of an outstanding 
even-odd effect that ties
the stability of a Majorana zero
mode to the parity 
of the hierarchical index $p$, as shown in Fig.~\ref{fig:plot-parafermion}.
In the next two Subsections we shall validate this even-odd effect by explicitly
showing that the parafermions of second hierarchical state ($p=3$)
possess a $\sqrt{2}$ contribution to their quantum dimension, similar to the primary
states ($p=1$) in Section~\ref{sec: primary states}; 
on the other hand, parafermions of third hierarchical state ($p=4$) repeat 
the same behavior as those of the first hierarchical state ($p=2$).

\subsection{Second Hierarchical State $(m,3)$}
\label{sec: level 2}

The low energy modes of the interface of the second hierarchical state are described
in terms of the fields
$\Phi^{T} = (\phi^{R}_{1},\phi^{R}_{2},\phi^{R}_{3},\phi^{L}_{1},\phi^{L}_{2},\phi^{L}_{3})$
and the Lagrangian Eq.~\ref{eq: hierarchical interface LL} has
\begin{equation}
\label{eq: interface K and Q level 2}
\begin{split}
&\,
\mathcal{K}_{m,3} = K_{m,3} \oplus (-K_{m,3})
\,,
\quad
Q_{m,3}^{T} = (q^{T}_{m,3},-q^{T}_{m,3}) 
\\
&\,
K_{m,3}=\left(
\begin{array}{ccc}
 2 m+1 & -1 & 0 \\
 -1 & 2 & -1 \\
 0 & -1 & 2 \\
\end{array}
\right)
\,,\quad
q_{m,3}^{T} = (1,0,0)
\,,
%\,,
\end{split}
\end{equation} 
where we adopt the hierarchical representation Eq.~\ref{eq: hierarchical K matrix - basis 1}.

The local charge neutral interactions
Eq.~\ref{eq: interactions and null vectors - charge conserving interface}
take the form
\begin{equation}
\begin{split}
U[\Lambda^{(0)}_{i}]
&\,=	
(\lambda_{i}/2) (\psi^{\dagger}_{i R}\psi_{i L} + \textrm{H.c.})
\\
&\,=
\cos{(\Lambda^{(0)}_{i} \mathcal{K}_{m,3}\Phi)}
\,,\quad
\textrm{for}~i = 1, 2, 3
\,,
\end{split}
\end{equation}
associated with the null vectors
\begin{equation}
\label{eq:null-charge-level 2}
\begin{split}
&\,
\Lambda^{(0)}_{1} = (1,0,0,1,0,0)
\\
&\,
\Lambda^{(0)}_{2} = (0,1,0,0,1,0)
\\
&\,
\Lambda^{(0)}_{3} = (0,0,1,0,0,1)
\,.
\end{split}
\end{equation}

Following Eq.~\ref{eq: integer vectors and interactions SC}, we now consider another set of null vectors 
\begin{equation}
\label{eq:null-sc-level 2}
\begin{split}
&\,
\Lambda^{}_{1} = (x_1,y_1,z_1,x_2,y_2,z_2)
\\
&\,
\Lambda^{}_{2} = \Lambda^{(0)}_{2} =  (0,1,0,0,1,0)
\\
&\,
\Lambda^{}_{3} = \Lambda^{(0)}_{3} =  (0,0,1,0,0,1)
\,,
%\,, i = 2~\textrm{and}~3,
\end{split}
\end{equation}
(where $x_{1}, ..., z_{2}$ are integers)
corresponding to the local interactions
\begin{equation}
U[\Lambda^{}_{i}] = \cos{(\Lambda^{}_{i}\mathcal{K}_{m,3}\Phi)}
\,\quad i = 1, 2, 3
\,,	
\end{equation}
where $U[\Lambda^{}_{2}]$ and $U[\Lambda^{}_{3}]$ are charge neutral
and $U[\Lambda^{}_{1}]$ carries charge $x_1 - x_2 \neq 0$. 
Imposing the null condition 
results in three equations
\begin{subequations}
\begin{equation}
\label{eq:null 1 p=3}
\begin{split}
&\,
\Lambda^{}_{1} \mathcal{K}_{m,3} \Lambda^{}_{1} = 
(2m+1) (x_{1}^2- x_{2}^2) - 2 x_{1} y_{1}
\\
&\,
+ 2 x_{2} y_{2} + 2 \left(y_{1}^2-y_{1} z_{1}-y_{2}^2+y_{2} z_{2}+z_{1}^2-z_{2}^2\right)
=0
\,,
\end{split}	
\end{equation}
\begin{equation}
\label{eq:null 2 p=3}
\Lambda^{}_{1}	\mathcal{K}_{m,3} \Lambda^{}_{2} = -x_1 + x_2 + 2(y_1 - y_2) - z_1 + z_2 = 0
\,,
\end{equation}
\begin{equation}
\label{eq:null 3 p=3}
\Lambda^{}_{1}	\mathcal{K}_{m,3} \Lambda^{}_{3} = -y_1 + y_2 + 2(z_1 - z_2) = 0
\,.
\end{equation}
\end{subequations}
Eqs.~\ref{eq:null 2 p=3} and~\ref{eq:null 3 p=3} result in
\begin{equation}
\label{eq:null 2 and 3 p=3 simplified}
z_1 - z_2 = \frac{x_1-x_2}{3}=\frac{y_1-y_2}{2}
\,,
\end{equation}
which gives
\begin{equation}
\label{eq:null 1 p=3 simplified}
\begin{split}
[2x_2 + 3(z_1 - z_2)](z_1-z_2) = 0
\,
\end{split}	
\end{equation}
upon substitution onto Eq.~\ref{eq:null 1 p=3}.

A non-trivial solution of Eq.~\ref{eq:null 2 and 3 p=3 simplified}
and Eq.~\ref{eq:null 1 p=3 simplified} yields the null vector
$
\Lambda^{}_{1} = (3t, y_2 + 4t, z_2 + 2t, -3t, y_2, z_2)
$ 
for $t \neq 0$ and $y_2, z_2 \in \mathbb{Z}$.
Furthermore, the non-zero minors
of $\mathcal{M}[\{ \Lambda_1, \Lambda_2, \Lambda_3 \}]$
belong in the set
$
\{\pm 2t, \pm 3 t, \pm 4t\}
$
from which the primitive condition follows for $t = \pm 1$.
Finally, setting $t=1$, $y_2 = -2$ and $z_2 = -1$, gives the null vector
\begin{equation}
\label{eq: Lambda 1 level 2}
\Lambda^{}_{1} = (3, 2, 1, -3, -2, -1)
\,,
\end{equation}
and the corresponding local interaction
\begin{equation}
\label{eq: third level interaction charge 6}
U[\Lambda_1] = \cos{\left(\Lambda^{}_{1} \mathcal{K}_{m,3} \Phi \right)} = 
\cos{\left[ (6m+1)(\phi^{R}_{1}+\phi^{L}_1) \right]}
\,.
\end{equation}

It follows from Eq.~\ref{eq: Lambda 1 level 2} and Eq.~\ref{eq: d2 general} that
\begin{equation}
N_{m,3} = \Lambda^{(0)}_{1}\,\mathcal{K}_{m,3}\,\Lambda_{1} = 2(6m+1)	
\,,
\end{equation}
which establishes the presence of $\mathbb{Z}_{2(6m+1)} \cong \mathbb{Z}_{2} \oplus \mathbb{Z}_{6m+1}$ parafermions 
\begin{equation}
\alpha_{i}\,\alpha_{j}
=
e^{i \frac{2\pi}{2(6m+1)}\textrm{sgn}(i-j)}
\alpha_{j}\,\alpha_{i}
\end{equation}
with quantum dimension
\begin{equation}
d_{m,3} = \sqrt{2}\times\sqrt{6m+1} = \sqrt{2}\times\mathcal{D}_{m,3}	
\,.
\end{equation}

This explicit calculation, therefore, confirms that the structure of the parafermions
in the second hierarchical state is similar to that observed
in the primary Laughlin states discussed in Section~\ref{sec: primary states},
with the quantum dimension of the parafermion being a manifestation
of both the bulk Abelian order and the non-trivial 1D superconductor.
Nevertheless, an important distinction emerges in this
case, for the interaction 
Eq.~\ref{eq: third level interaction charge 6}
represents a 
condensate of 
charge $\mathcal{Q}[\Lambda^{}_{1}] = Q_{m,3}\Lambda^{}_{1} = 6$.
The stability of parafermions,
as seen in 
Eq.~\ref{eq:parafermions},
is captured by 
the expectation value of the 
charge $\frac{3}{6m+1}$
operator
\begin{equation}
\langle\, \mathcal{O}_{\frac{3}{6m+1}}  \,\rangle
\equiv
\langle\, e^{i\frac{\Lambda^{(0)}_{1}\,\mathcal{K}_{m,3}\,\Lambda_{1}}{2(6m+1)}}	\,\rangle
=
\langle\, e^{i\frac{(\phi^{L}_{1}+\phi^{R}_{1})}{2}} \,\rangle
\neq
0
\,,
\end{equation}
on the segments of the interface gapped by this interaction.  
As an example, the interface between two FQH states at filling fraction
$\nu=3/7$ (corresponding to $m=1$, $p=3$), can give rise to $\mathbb{Z}_{14} \cong \mathbb{Z}_{2} \oplus \mathbb{Z}_{7}$ 
parafermions along the domain walls described here. 
The charge $6$ condensate, in this case, is formed by condensation of fractional charge $3/7$.

It is instructive to seek an understanding
of this charge $6$ gapped interface in the 
the representation Eq.~\ref{eq: hierarchical Chern-Simons - basis 2},
\begin{equation}
\label{eq: K matrix second hierarchical basis 2}
\begin{split}
&\,
\tilde{K}_{m,3}
=
\begin{pmatrix}
2m+1 & 2m & 2m
\\
2m & 2m+1 & 2m
\\
2m & 2m & 2m+1
\end{pmatrix}
\,,
~
\tilde{q}_{m,3} = (1,1,1)
\end{split}	
\end{equation}
where the $K$ matrix and charge vectors
resemble a tri-layer FQH state, where each layer
carries unit charge.
The SL($3,\mathbb{Z}$) transformation
to this new basis
\begin{equation}
W_{3} = 
\begin{pmatrix}
1 & 0 & 0
\\
1 & 1 & 0
\\
1 & 1 & 1
\end{pmatrix}	
\end{equation}
changes the null vector to
\begin{equation}
\begin{split}
\tilde{\Lambda}_{1} 
&\,= 
\left[ \left( W^{-1}_{3}\right)^{T} \oplus \left( W^{-1}_{3}\right)^{T} \right]  \,\Lambda_{1}
\\
&\,=
(1,1,1,-1,-1,-1)^{T}
\,.
\end{split}
\end{equation}
Then the interaction 
\begin{equation}
\label{eq: interaction tilde basis second hierarchical state}
\begin{split}
U[\tilde{\Lambda}_{1}]
&\,=
\tilde{\lambda}\cos{\left( \tilde{\Lambda}_{1} \tilde{\mathcal{K}}_{m,3} \tilde{\Phi} \right)}
\\
&\,
\sim
\frac{\tilde{\lambda}}{2}
\,
\tilde{\psi}_{1R}
\tilde{\psi}_{2R}
\tilde{\psi}_{3R}
\tilde{\psi}_{1L}
\tilde{\psi}_{2L}
\tilde{\psi}_{3L}	
+
\textrm{H.c.}
\end{split}
\end{equation}
is manifestly a charge $6$ operator involving
pairing of $3$ local fermions on each side of the interface.
This interaction is a generalization
of the charge $2$ pairing at the interface
of Laughlin states.

\begin{figure}
\includegraphics[width=\columnwidth]{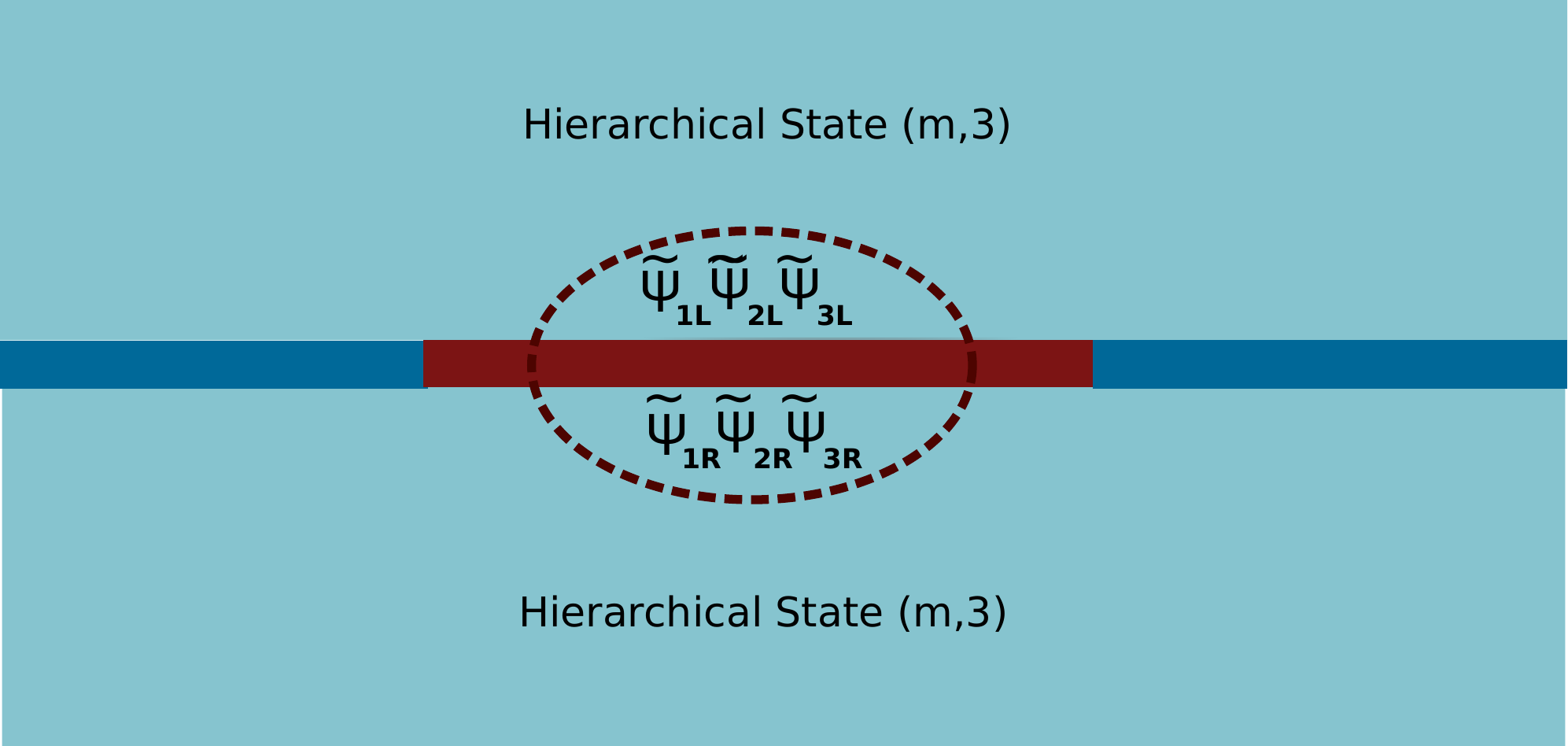}
\caption
{
Interface between the second hierarchical FQH state $(m,3)$.
The red segment represents the interface
gapped by the charge $6$ interaction   
Eq.~\ref{eq: interaction tilde basis second hierarchical state}.	
}
\label{fig:odd-p-condensate}
\end{figure}

Similarly to the discussion 
of the first hierarchical state,
we compare our set up
with the theory of extrinsic defects associated
with anyonic symmetries of the Abelian phase.
In Ref.~\cite{BarkeshliJianQi-2013-a}, the $\mathbb{Z}_3$
twisted defects associated with the layer permutations 
of the Abelian phase characterized by the K
matrix
\begin{equation}
\label{eq:K matrix trilayer m l l}
K_{m,\ell,\ell}
=
\begin{pmatrix}
m & \ell & \ell
\\
\ell & m & \ell
\\
\ell & \ell & m
\end{pmatrix}
\,	
\end{equation}
where studied and it was shown that domain walls separating 
two distinct gapping charge conserving gap terms support parafermions
with quantum dimension $d_{\mathbb{Z}_3} = | m - \ell |$.
Comparing Eq.~\ref{eq: K matrix second hierarchical basis 2}
and
Eq.~\ref{eq:K matrix trilayer m l l}
shows that that $\mathbb{Z}_3$ twist
defect of the second hierarchical state
is trivial, since $d_{\mathbb{Z}_3}=1$.
Furthermore, it can be shown~\cite{comment-chargeconjugation} that
twist defects associated with charge conjugation
anyonic symmetry are associated with Majorana fermions,
in contrast with the $\mathbb{Z}_{2(6m+1)}$ parafermions discussed here.

\subsection{Third hierarchical state $(m,4)$}
\label{sec: level 3}

The interface of the third hierarchical FQH state $(m,4)$ with filling fraction $\nu_{m,4} = \frac{4}{8m+1}$ 
supports the low energy	mode fields
$
(\phi^{R}_{1},\phi^{R}_{2},\phi^{R}_{3},\phi^{R}_{4},\phi^{L}_{1},\phi^{L}_{2},\phi^{L}_{3},\phi^{L}_{4})
$
where the Lagrangian Eq.~\ref{eq: hierarchical interface LL} has
\begin{equation}
\label{eq: interface K and Q level 3}
\begin{split}
&\,
\mathcal{K}_{m,4} = K_{m,4} \oplus (-K_{m,4})
\,,
\quad
Q_{m,4}^{T} = (q^{T}_{m,4},-q^{T}_{m,4}) 
\\
&\,
K_{m,4} 
= 
\begin{pmatrix}
2m+1 & -1 & 0 & 0
\\
-1 & 2 & -1 & 0
\\
0 & -1 & 2 & -1
\\
0 & 0 & -1 & 2
\end{pmatrix}
\,,\quad
q_{m,4}^{T} = (1,0,0,0)
\,.
%\,,
\end{split}
\end{equation} 

Charge neutral null vectors parametrizing the 
gap openining interactions Eq.~\ref{eq: interactions charge conserving interface}
read
\begin{equation}
\label{eq:null-charge-level 3}
\begin{split}
&\,
\Lambda^{(0)}_{1} = (1,0,0,0,1,0,0,0)^{T}
\\
&\,
\Lambda^{(0)}_{2} = (0,1,0,0,0,1,0,0)^{T}
\\
&\,
\Lambda^{(0)}_{3} = (0,0,1,0,0,0,1,0)^{T}
\\
&\, 
\Lambda^{(0)}_{4} = (0,0,0,1,0,0,0,1)^{T}
\,.
\end{split}
\end{equation}

Carrying out an analysis similar 
to that discussed in Sections~\ref{sec: level 1},~\ref{sec: level 2} and~\ref{sec: level 3}, 
we find
that the null integer vectors of the charge non-conserving gapped interface are
\begin{equation}
\label{eq:null-sc-level 3}
\begin{split}
&\,
\Lambda_{i}
=
\begin{cases}
(2,3,2,1,-2,0,0,0)^{T} & i = 1
\\
\Lambda^{(0)}_{i} & i = 2, 3, 4
\end{cases}
\end{split}
\end{equation}
Notice that the interaction
\begin{equation}
\begin{split}
U[\Lambda^{}_{1}] &\,= \cos{(\Lambda^{}_{1} \mathcal{K}_{m,p}\Phi)}
\\
&\,\sim
\psi^{2}_{1R}
\,
\left(
\psi^{3}_{2R}
\psi^{2}_{3R}
\psi^{}_{4R}
\right)
\,
\psi^{2}_{1L}
+
\textrm{H.c}
\end{split}	
\end{equation}
is a charge $4$ cluster operator
where $\psi_{1R}$ and $\psi_{1L}$
are charge $1$ local operators with fermionic statistics
and $\psi_{2R},\psi_{3R}, \psi_{4R}$ are charge zero
local operators with bosonic statistics.

From Eq.~\ref{eq:null-sc-level 3} and Eq.~\ref{eq: d2 general} we get
\begin{equation}
N_{m,4} = \Lambda^{(0)}_{1}\,\mathcal{K}_{m,4}\,\Lambda_{1} = (8m+1)	
\,,
\end{equation}
which establishes the presence of $\mathbb{Z}_{8m+1}$ parafermions 
\begin{equation}
\alpha_{i}\,\alpha_{j}
=
e^{i \frac{2\pi}{8m+1}\textrm{sgn}(i-j)}
\alpha_{j}\,\alpha_{i}
\end{equation}
with quantum dimension
\begin{equation}
d_{m,4} = \sqrt{8m+1} = \mathcal{D}_{m,4}	
\,.
\end{equation}

This result shows, that the parafermions in the third hierarchical state 
behave similarly to the first hierarchical state discussed in Section~\ref{sec: level 1}. 
This non-trivial condensate is manifested in the expectation 
value of of the charge $\frac{4}{8m+1}$ operator
\begin{equation}
\Big\langle\,
\mathcal{O}_{\frac{4}{8m+1}}
\,
\Big\rangle
\equiv
\Big\langle\,
e^{i\frac{\Lambda^{(0)}_{1}\,\mathcal{K}_{m,4}\,\Lambda_{1}}{8m+1}}
\,
\Big\rangle
\neq 0
\,.
\end{equation}
So, for instance, 
the $\nu=4/9$ FQH state (m = 1, p = 4)
is seen to support 
$\mathbb{Z}_{9}$ 
parafermions along domain walls at its gapped edge. 
The charge $4$ condensate is formed by a cluster of 
$9$ quasiparticles with charge $4/9$.

\section{Hierarchy of parafermions: General Case}
\label{sec:General parafermion hierarchy}

The properties of the parafermions zero modes stabilized
on domain walls of the $p=1,2,3,4$ hierarchical states
discussed in Section.~\ref{sec: hierarchy examples} reveal
a remarkable dependence on the parity of the index $p$, which labels the depth of the 
hierarchy. 
This dependence reflects an interplay between the bulk topological order,
which gives rise to quasiparticle fractionalization, and 
the 1D SPT order that stabilizes Majorana zero modes in non-trivial topological superconductors. 
An appealing mechanism to account
for such an even-odd dependence emerges when 
upon expressing the bulk topological order of the hierarchical
state in the representation Eq.~\ref{eq: hierarchical Chern-Simons - basis 2}, 
where the K-matrix gives an interpretation
of the bulk topological order as a series of 
$p$ Laughlin-type layers (as indicated in the diagonal odd integers
$2m+1$) coupled to each other by bosonic correlations (indicated
by the off-diagonal even integers $2m$.)
As such, 
the number of pairs of counter-propagating fermion modes
in interfaces of hierarchical states $(m,p)$ and $(m,p+2)$
differ by $2$. In the limit, where these Laughlin FQH layers
are decoupled (which would correspond to $K = (2m+1)\,\textrm{diag}(1,...,1)$),
the even-odd effect associated with the stability of Majorana zero modes
is a direct consequence of the $\mathbb{Z}_2$ stability (instability) associated
with an odd (even) number of Majorana zero modes per domain. 
Remarkably, our analysis will show that this structure
persists even when the layers are coupled, according to the K-matrix Eq.~\ref{eq: hierarchical K matrix - basis 2}.

One of the goals of this Section is to show that this $\mathbb{Z}_2$ 
pattern indeed persists \textit{for all} hierarchical states, whose bulk topological order are represented
by Eq.~\ref{eq: hierarchical Chern-Simons - basis 1} or, equivalently, Eq.~\ref{eq: hierarchical Chern-Simons - basis 2}. 
We will show that the interactions require breaking of charge conservation
in such a way that the charge of the condensate depends on the hierarchical level $p$.
Notably, while the primary Laughlin ($p=1$) admit a charge 2
condensate that, in principle, can be induced by a weak-pairing mechanism (or by
proximity to a superconductor), for generic states of the hierarchical sequence,
the $\mathcal{Q}_{m,p} > 2$ charge of the condensate signals that a non-BCS strong coupling mechanism
is at play. This situation departs significantly
from the stability of Majorana zero modes in superconducting wires~\cite{Kitaev-2001}
as well as in interfaces of Laughlin-type states.~\cite{Lindner-2012,Clarke-2013,Cheng-2012,Vaezi-2013}

An interesting property of the condensate
is that it involves clustering of $2mp+1$
quasiparticles of charge $\nu_{m,p} = \frac{p}{2mp+1}$.
This scenario of parafermions being stabilized by clustering 
of quasiparticles is analogous to 
clustering property of non-Abelian Read-Rezayi FQH states,~\cite{read-rezayi-1999}
where electrons (or composite fermions) forming an order-$k$ cluster,
give rise to an incompressible state that supports non-Abelian bulk excitations
and chiral charge neutral $\mathbb{Z}_k$ parafermions on the boundary. (The special case $k=2$
corresponds to the Moore-Read states with a chiral Majorana fermions at the boundary,
which is a candidate topological order for the $\nu=5/2$ FQH.) 

In Sections~\ref{sub:hierarchical_states_p_odd_} 
and~\ref{sub:hieracrchical_states_p_even} we discuss, respectively, the $p=$ odd
and $p=$ even hierarchical states in generality, where we shall
provide explicit expressions for the null vectors and, consequently, the local interactions
that gap the interface and give rise to parafermion zero modes
localized on domain walls.

\subsection{Hierarchical states: $p=\textrm{odd integer}$} % (fold)
\label{sub:hierarchical_states_p_odd_}

Consider the interface between hierarchical FQH states 
with filling fraction 
\begin{equation}
\label{eq:filling odd p}
\begin{split}
\nu_{m,p} = \frac{p}{2mp+1} 
\,,~~
m \in \mathbb{Z}_{+}
\,,~ 
p=1,3,5,...
\,. 		
	\end{split}	
\end{equation}

As discussed in
Section~\ref{sec:Gapped interfaces in Hierarchical FQH states},
this interface admits a gapped region realized by the local charge neutral interactions
Eq.~\ref{eq: interactions and null vectors - charge conserving interface}. 
We now demonstrate that the interface formed by the states in Eq.~\ref{eq:filling odd p}
admits a set of local gap opening interactions
that breaks $U(1)$ charge conservation and 
represent a charge $\mathcal{Q}_{m,p}$
condensate where
\begin{equation}
\label{eq:charge of condensate odd p}
\mathcal{Q}_{m,p} = 2p
\,,~~
m \in \mathbb{Z}_{+}
\,,~ 
p=1,3,5,...	
\end{equation}
In the representation 
Eq.~\ref{eq: hierarchical Chern-Simons - basis 1},
this condensate
is realized by the local interactions
\begin{equation}
\label{eq:sc interactions - general odd p}
U[\Lambda^{}_{i}] = \cos{(\Lambda^{}_{i} \mathcal{K}_{m,p}\Phi)}
\,,~~ i = 1, ... , p
\,,
\end{equation}
represented by the integer vectors
\begin{equation}
\label{eq:integer-vecs-sc p odd}
\begin{split}
&\,
\Lambda^{}_{1} =
\begin{pmatrix}
v_{p}
\\
-v_{p}
\end{pmatrix}
\,,\quad
v_{p}
=
\begin{pmatrix}
p
\\
p-1
\\
\vdots
\\
1
\end{pmatrix}
\\
&\,
\Lambda^{}_{i} = \Lambda^{(0)}_{i} =
\begin{pmatrix}
e_i
\\
e_i
\end{pmatrix}
\,,\quad i = 2, ..., p	
\,,
\end{split}
\end{equation}
which satisfy the null condition
\begin{equation}
\label{eq:null condition sc vecs p odd}
\Lambda^{}_{i} \mathcal{K}_{m,p} \Lambda^{}_{j}	
\,,\quad i, j = 1, ..., p
\,.
\end{equation}

It follows immediately from Eq.~\ref{eq:integer-vecs-sc p odd}
that the interaction corresponds to a condensate of charge 
$\mathcal{Q}[\Lambda^{}_{1}] = \Lambda^{}_{1} Q_{m,p} = 2p$,
in accordance with Eq.~\ref{eq:charge of condensate odd p}.
(Note that for $2 \leq i \leq p$ : $\mathcal{Q}[\Lambda^{}_{i}] = \Lambda^{}_{i} Q_{m,p} = 0$.)
Moreover, the non-zero minors of $\mathcal{M}[\{ \Lambda \}]$
can be shown to form the set
$\{ \pm (p-1), \pm p, \pm (p+1) \}$
whose greatest common divisor is one,
which shows that the integer vectors
in Eq.~\ref{eq:integer-vecs-sc p odd} are primitive.
Furthermore, to establish the null condition 
Eq.~\ref{eq:null condition sc vecs p odd},
we first realize that this condition is clearly satisfied for $i,j = 2,...,p$
as a consequence of Eq.~\ref{eq:null condition charge conserving vecs}.
Then, the remaining non-trivial conditions we need to 
show are for $i=1$ and $j=1,...,p$. 
To establish this result, all we need is the identity
\begin{equation}
\label{eq: K times v odd p}
K_{m,p}\,v_{p} = 
\begin{pmatrix}
2mp+1
\\
0\\
\vdots
\\
0
\end{pmatrix} 
=
(2mp+1)e_1
\,,
\end{equation}
where $v_p$ is $p$ dimensional integer vector
defined in 
Eq.~\ref{eq:integer-vecs-sc p odd}.
To demonstrate this result, let $K_{m,p}\,v_{p} = \sum^{p}_{k = 1} a_{k}\,e_{k}$.
Except for the first and last rows, the remaining rows of $K_{m,p}$ are 
formed by consecutive entries $-1,2,1$ and the remaining ones equal to zero.
The first row has $(K_{m,p})_{11} = 2m+1$ and $(K_{m,p})_{12} = -1$.
The last row has $(K_{m,p})_{p, p-1} = -1$ and $(K_{m,p})_{p,p} = 2$. Putting all together,
\begin{equation}
\begin{split}
&\,
a_{1} = p(2m+1) + (-1)\times(p-1) = 2mp+1
\\
&\,
a_{k} = (-1)\times(k+1) + 2\times(k) + (-1)\times(k-1)
\\
&\, ~~~ = 0\,, ~~ 2 \leq k \leq p-1
\\
&\,
a_{p} = (-1)\times(2) + 2\times(1) = 0
\,,		
	\end{split}	
\end{equation}
which proves
Eq.~\ref{eq: K times v odd p}.
Finally, by taking into 
account the orthonormal basis vectors $e_i$,
it is straightforward to
verify the null condition
Eq.~\ref{eq:null condition sc vecs p odd}.

The form of the charge $2p$ interaction,
which follows directly from 
Eq.~\ref{eq: K times v odd p}, is
\begin{equation}
U[\Lambda_1] = \cos{\left( \Lambda_1\,\mathcal{K}_{m,p}\,\Phi \right) }
=
\cos{\left[ (2mp+1) (\phi^{R}_{1} + \phi^{L}_{1}) \right]}	
\end{equation}

In the basis given by  
Eq.~\ref{eq: hierarchical Chern-Simons - basis 2}
the integer vectors $\Lambda_1$ transforms to
\begin{equation}
\tilde{\Lambda}_1 = \left(W^{-1}\right)^{T}\,\Lambda_1
=
(1,1,...,1,-1,-1,...,-1)^{T}
\,.
\end{equation}
The meaning of the charge $2p$ interaction in this
representation is manifestly given by
\begin{equation}
\label{eq: general interaction charge 2p}
\begin{split}
U[\tilde{\Lambda}_1]
&\,=
\cos{\left( \tilde{\Lambda}_1\,\tilde{\mathcal{K}}_{m,p}\,\tilde{\Phi} \right) }
\\
&\,
\sim
\left(\tilde{\psi}^{L}_{1}\tilde{\psi}^{R}_{1}\right)
...
\left(\tilde{\psi}^{L}_{p}\tilde{\psi}^{R}_{p}\right)
+
\textrm{H.c.}
\end{split}
\end{equation}
which represents the cluster of $2p$ fermions, each one carrying charge $q=1$.
Eq.~\ref{eq: general interaction charge 2p}
generalizes, to every odd value of $p$, the 
charge $6$ interaction that gaps the interface of the hierarchical
state $(m,3)$ depicted in 
Fig.~\ref{fig:odd-p-condensate}.

Finally, the quantum dimension of the parafermion localized at the domain wall
between the segments of the interface that are gapped by interactions 
Eqs.~\ref{eq: interactions charge conserving interface}
and 
Eq.~\ref{eq:sc interactions - general odd p} follows from 
\begin{equation}
\begin{split}
N_{m,p}
=
\Lambda^{(0)}_{1}\mathcal{K}_{m,p}\Lambda_{1}	
&\,=
\begin{pmatrix}
e^{T}_{1} & e^{T}_{1}
\end{pmatrix}	
\begin{pmatrix}
(2mp+1)e_1
\\
(2mp+1)e_1
\end{pmatrix}
\\
&\,= 2(2mp+1)
\,,
\end{split}
\end{equation}
which establishes
the existence of parafermions of quantum dimension
\begin{equation}
d_{m,p} = \sqrt{2} \times \sqrt{2mp+1}
\,,
\end{equation}
at the 
interface of hierarchical states 
filling fraction
$
\nu_{m,p} = \frac{p}{2mp+1} 
\,,~~
m \in \mathbb{Z}_{+}
\,,~ 
p=1,3,5,...
$.

% subsection hierarchical_states_p_odd_ (end) 

\subsection{Hierarchical states: $p = \textrm{even integer}$} % (fold)
\label{sub:hieracrchical_states_p_even}

Consider the interface between hierarchical FQH states 
with filling fraction 
\begin{equation}
\label{eq:filling even p}
\begin{split}
\nu_{m,p} = \frac{p}{2mp+1} 
\,,~~
m \in \mathbb{Z}_{+}
\,,~ 
p=2,4,6,...
\,. 		
	\end{split}	
\end{equation}

As discussed in
Section~\ref{sec:Gapped interfaces in Hierarchical FQH states},
this interface admits a gapped region realized by the local charge neutral interactions
Eq.~\ref{eq: interactions and null vectors - charge conserving interface}. 
We now prove that the interface formed by the states in Eq.~\ref{eq:filling even p}
can be gapped by local interactions that break charge conservation
symmetry and give rise to a condensate of charge 
\begin{equation}
\label{eq:charge of condensate even p}
\mathcal{Q}_{m,p} = p
\,,~~
m \in \mathbb{Z}_{+}
\,,~ 
p=2,4,6,...
\end{equation}

In the representation 
Eq.~\ref{eq: hierarchical Chern-Simons - basis 1},
this charge $p$ condensate
is realized by the local interactions
\begin{equation}
\label{eq:sc interactions - general even p}
U[\Lambda^{}_{i}] = \cos{(\Lambda^{}_{i} \mathcal{K}_{m,p}\Phi)}
\end{equation}
where the integer vectors read
\begin{equation}
\label{eq:integer-vecs-sc p even}
\Lambda_{i} = 
\begin{cases}
      (\frac{p}{2},p-1,p-2,~...~,1,-\frac{p}{2},0,0,~...~,0)^{T} & $i=1$\\
      \Lambda^{(0)}_{i} & i = 2,...,p\\
    \end{cases}
\end{equation}
%\end{widetext}
and
satisfy the null condition
\begin{equation}
\label{eq: null condition even p}
\Lambda_i \, \mathcal{K}_{m,p}\,\Lambda_j = 
0
\,,~~ \forall i, j = 1, ... , p
\,.	
\end{equation}

With the null vectors
Eq.~\ref{eq:integer-vecs-sc p even}
we directly find that this corresponds to a 
condensate of charge
$\mathcal{Q}[\Lambda^{}_{1}] = \Lambda^{}_{1} Q_{m,p} = p$,
as given by
Eq.~\ref{eq:charge of condensate even p}.
(Notice that for $2 \leq i \leq p$ : $\mathcal{Q}[\Lambda^{}_{i}] = \Lambda^{}_{i} Q_{m,p} = 0$.)
Moreover, from  the explicit form of the 
integer vector $\Lambda_1$ in 
Eq.~\ref{eq:integer-vecs-sc p even},
one verifies that 
the charge $p$ interaction
\begin{equation}
\begin{split}
U[\Lambda^{}_{1}] &\,= \cos{(\Lambda^{}_{1} \mathcal{K}_{m,p}\Phi)}
\\
&\,\sim
\psi^{p/2}_{1R}
\,
\left(
\psi^{p-1}_{2R}
\psi^{p-2}_{3R}
...
\psi^{2}_{(p-1)R}
\psi^{}_{pR}
\right)
\,
\psi^{p/2}_{1L}
+
\textrm{H.c}
\end{split}	
\end{equation}
represents a cluster operator
where $\psi_{1R}$ and $\psi_{1L}$
are charge $1$ local operators with fermionic statistics
and $\psi_{2R}, ..., \psi_{p R}$ are charge zero
local operators with bosonic statistics.

We now demonstrate the validity of
Eq.~\ref{eq: null condition even p}.
Since $\{ \Lambda^{(0)}_{i}\}$, $i = 2, ..., p$ forms, by construction, a subset 
of null vectors, the only non-trivial relations left to be verified are
$
\Lambda_{i}\mathcal{K}_{m,p}\Lambda_{1} = 0
$
for $i = 1, ..., p$. In order to establish these conditions,
we directly calculate 
\begin{equation}
\label{eq: KL1 for p even}
\begin{split}
\mathcal{K}_{m,p}
&\,
\Lambda_{1}
=
\begin{pmatrix}
\frac{p}{2} (2m-1) + 1
\\
\frac{p}{2}
\\
0
\\
\vdots
\\
0
\\
0
\\
\frac{p}{2}(2m+1)
\\
-\frac{p}{2}
\\
0
\\
\vdots
\\
0
\end{pmatrix}	
\,,	
\end{split}	
\end{equation}
leading to  
\begin{equation}
\begin{split}
&\,
\Lambda_{1}\mathcal{K}_{m,p}\Lambda_1
=
\\
&\,
\frac{p}{2}
\left[  
\frac{p}{2} (2m-1) + 1
\right]	
+
(p-1)\frac{p}{2}
-
\frac{p}{2}
\left[
\frac{p}{2}(2m+1)
\right]
=0
\\
&\,
\Lambda_{2}\mathcal{K}_{m,p}\Lambda_1
=
\frac{p}{2} - \frac{p}{2} = 0
\\
&\,
\Lambda_{i}\mathcal{K}_{m,p}\Lambda_1
=
0
\,,
~~ i = 3, ..., p
\,,
% ~~ (\textrm{since}~e^{T}_{i}\cdot e_{1,2} = 0)
\end{split}
\end{equation}
where the last equation 
is a consequence of 
$
e^{T}_{i}\cdot e_{1,2} = 0
$
for 
$
i = 3, ..., p
$.
This then establishes the null condition 
Eq.~\ref{eq: null condition even p}.

Finally, the quantum dimension of parafermion localized at the domain wall
between the segments of the interface that are gapped by interactions 
Eqs.~\ref{eq: interactions charge conserving interface}
and 
Eq.~\ref{eq:sc interactions - general even p} follows from 
\begin{equation}
\begin{split}
N_{m,p}
=
\Lambda^{(0)}_{1}\mathcal{K}_{m,p}\Lambda_{1}	
&\,=
2mp+1
\,,
\end{split}
\end{equation}
which shows establishes
the existence of parafermions
with quantum dimension
\begin{equation}
d_{m,p} = \sqrt{2mp+1}
\,
\end{equation}
at the interface with 
filling fraction
$
\nu_{m,p}
=
\frac{p}{2mp+1}
$,
with
$
m \in \mathbb{Z}_{+}
$
and
$p = 2, 4, 6, ...$

% subsubsection hieracrchical_states_p_even (end)

We have then explicitly demonstrated,
in Sections~\ref{sub:hierarchical_states_p_odd_}
and~\ref{sub:hieracrchical_states_p_even},
the existence of a local charge condensate,
Eq.~\ref{eq:condensate Q - general},
that stabilizes 
non-Abelian parafermions with quantum dimensions
given by
Eq.~\ref{eq: d and e relation}
and depicted in 
Fig.~\ref{fig:plot-parafermion}.

\section{Summary and Outlook} % (fold)
\label{sec:summary_and_discussions}

In this work, we have established a correspondence
between the sequence of Abelian hierarchical FQH states in the first Landau level 
and a class of extrinsic non-Abelian zero modes localized on domain walls
that separate charge neutral and $U(1)$ symmetry broken gapped segments of the interfaces.
Our analysis of the low energy properties
of the bulk hierarchical state employed
the hydrodynamical Chern-Simons theory
by which 
the FQH system with Hall conductance $\sigma_{xy}(m,p) = \frac{e^2}{h}\frac{p}{2mp+1}$
is represented in terms of a $p$-component $U(1)$ Chern-Simons gauge theory
parametrized by an integer valued 
$K$ matrix.~\cite{Blok-Wen-1990,Read-1990,FrohlichZee-1991,WenZee-1992,FrohlichThiran-1994}
The edge of such hierarchical state, in turn, supports $p$
chiral low energy modes described by a $p$-component  
chiral boson field whose commutation relations are determined by the K matrix.
As such, we have studied gap opening processes
in a homogeneous interface with $p$ pairs of counter-propagating
modes, as depicted in Fig.~\ref{fig:domains-interface}.
Gapping these modes at the interface is realized by $p$ local
sine-Gordon type operators.

Through a detailed examination of the locality and frustration
free conditions of sine-Gordon operators on the homogeneous
interface, we have found that the hierarchical states admit
$U(1)$ symmetry breaking interactions that give rise
to a condensate whose charge is a function of the hierarchical
index $p$, as per Eq.~\ref{eq:condensate Q - general}.
Therefore, our results show 
that a charge $2$ condensate only occurs 
for $p=1$ and $p=2$,
i.e.,
for the primary Laughlin states
with filling fraction $\frac{1}{2m+1}$ and the first hierarchical states 
with filling fraction $\frac{2}{4m+1}$,
for integer $m > 1$.
(We note that charge 2 condensates formed the basis of earlier
studies of parafermions in interfaces and trenches of 
Laughlin states~\cite{Lindner-2012,Clarke-2013,Cheng-2012,Vaezi-2013}
and the particle-hole conjugate of the $\nu=1/3$ Laughlin state
at filling $\nu=2/3$.~\cite{Mong-2014})
For the general $p > 2$ case investigated in this work, on the other hand, 
we have found that
the local $U(1)$ symmetry breaking interactions 
involve a non-conventional charge clustering mechanism 
whereby more than two electrons are glued together.
Our findings then open the interesting possibility of exploring these gapped interfaces 
as a basis for constructing families of unconventional U(1) symmetry broken phases in 2D 
by promoting the interfaces to a ``wire network",
in the spirit of Ref.~\cite{teo-kane-2014}.

One of the main results of this work
was establishing that the properties of
parafermions zero modes stem from the 
existence of a cluster state
of charge given by Eq.~\ref{eq:condensate Q - general},
which translates into a cluster of fractionalized quasiparticles of charge $p/(2mp+1)$.
This state bears a striking resemblance with the Read-Rezayi FQH
states that represent non-Abelian
FQH states where electrons form cluster states.
In fact, this correspondence has been
explored in Ref.~\cite{Mong-2014},
where it was shown that superconducting islands in the 
$\nu=2/3$ FQH state harbor $\mathbb{Z}_3$ parafermions
that are closely related to the neutral parafermion
excitations of the $\mathbb{Z}_3$ Read-Rezayi FQH,
whose ground state wavefunction encodes clustering
of $3$ electrons. 
From this perspective, the results obtained here
for the entire sequence of Abelian hierarchical FQH states
establish a rich connection between two distinct families of 
Abelian and non-Abelian topological orders,
and suggest, in particular, a route to describe the hierarchy of 
non-Abelian phases~\cite{lan-wen-2017} via the deconfinement of 
extrinsic parafermion zero modes in corresponding Abelian phases of matter,
which is an important topic worth of further investigation.
Furthermore, since the parafermions
in the setting considered here do not manifest any direct relationship
with anyonic symmetries,
the deconfinement of these non-Abelian defects
may require a theoretical treatment that differs
from those of  
Refs.~\cite{BBCW-2014} 
and~\cite{TeoHughesFradkin-2015},
which dealt with the deconfinement of twist defects related to
symmetries of the anyon group. 

We have found an appealing dependence of the charge condensate and the 
quantum dimension of parafermions on the parity of the level hierarchy $p$,
as shown in Fig.~\ref{fig:plot-parafermion}.
Borrowing from insights 
related to the multi-layer representation
of the  
K matrix of the hierarchical state -- despite the fact that each hierarchical state
we studied is understood to be realized in a monolayer
system in the lowest Landau level --, 
we have argued that the even-odd dependence on $p$ 
is indicative of the $\mathbb{Z}_2$ stability 
of Majorana zero modes in 1D topological superconductors,
where the parity of $p$ matches the parity of Majorana zero models per domain, 
according to the quantum dimension 
Eq.~\ref{eq: d and e relation}.
According to this result, the quantum dimension of the parafermions depends both on the 
bulk Abelian topological order via the total quantum dimension $\mathcal{D}_{m,p}$ 
of the hierarchical Abelian bulk phase,
as well as on the Majorana modes stabilized by conservation
of fermion parity
in
1D gapped fermionic phases of matter.~\cite{qi-hughes-zhang-2008,schnyder-2009,kitaev-2009}

We close by pointing to a relation between 
the non-zero charge condensate at the interface 
of hierarchical FQH states and the entanglement
entropy associated with an entanglement cut across the 
interface.~\cite{QiKatsuraLudwig-2012,Lundgren-2013,ChenFradkin_2013,FurukawaShunsukeKim-2011}
In that regard, it is possible to show that 
the interactions giving rise to the charge condensate 
in the hierarchical states
are invariant under a global $\mathbb{Z}_{k} \times \mathbb{Z}_k$ 
symmetry [where $k=p$ ($k=p/2$) for odd (even) values of $p$],
which correspond to transformations on the local operators 
having support on each side of the interface.
It was shown in Ref.~\cite{scmh-2018}
that the existence of 
local gap opening interactions possessing
such discrete symmetry gives rise to a non-trivial 
$\log{(k)}$ correction to the bulk universal value of the 
topological entanglement entropy, which, in 
non-chiral bulk Abelian phases, characterize
the onset of a 1D gapped SPT chain along the interface.
(See also 
Ref.~\cite{CanoMulliganHughes-2015} 
for an early discussion of entanglement corrections 
in 2D Abelian phases of matter and 
Ref.~\cite{Williamson-2019}
for a relationship between such entanglement corrections
and string order parameters.)
We stress, however, that hierarchical states studied here
are chiral phases, which implies that the parafermion
zero modes are not protected by the emergent $\mathbb{Z}_{k} \times \mathbb{Z}_k$
symmetry of the local interactions that stabilize the condensate.
Nevertheless, we note that the local interactions 
we have discussed here by no means exhaust the possible
classes of gapped interfaces that can be formed
in hierarchical states. 
It is then an important open question whether such hierarchical
interfaces can support also genuine 1D SPT phases of matter
protected by other classes of discrete symmetries,
as recently discussed in Ref.~\cite{scmh-2018}. 

In summary, leveraging on the anyon condensation mechanism that gives rises 
to the hierarchical sequence of Abelian FQH states, we have 
discovered a remarkably rich sequence of non-Abelian parafermions 
that are stabilized by clustered states of electrons and quasiparticles
on their interfaces. Our study opens an exciting possibility
to gain a deeper understanding of the structure of 2D non-Abelian states
by exploring more familiar and well understood Abelian phases of matter.

\section*{Acknowledgments} % (fold)
%\label{sec:acknowledgments}

I wish to thank Jason Alicea for an insightful discussion and 
Paul Fendley for a feedback on the manuscript.
L.H.S. is supported by a faculty startup at Emory University.
% section acknowledgments (end)

\bibliography{parafermion}

\begin{thebibliography}{66}
\bibitem [{com({\natexlab{b}})}]{comment-chargeconjugation}%
  \BibitemOpen
  \href@noop {} {} ({\natexlab{b}}),\ \bibinfo {note} {{Let there be a segment
  of the interface gapped by local charge neutral interactions parametrized by
  the null vectors $\{ \Lambda^{(0)}_{i} = (e_i, e_i)~|~ i = 1, ... , p \}$ and
  another one gapped by the charge $2$ interactions with $\{ \Lambda^{(2)}_{i}
  = (e_i, -e_i)~|~ i = 1, ... , p \}$. Analysis of the commutation relations of
  the operators similar to Eq.~\ref{eq:gammas and N} shows that the quantum
  dimension of the zero mode is given the g.c.d. of the set of integer numbers
  $\{ \Lambda^{(0)}_{i} \mathcal{K}_{m,p} \Lambda^{(2)}_{j} | i,j = 1, ...,
  p\}$. From the structure of the K matrix of the hierarchical state, we find
  that the g.c.d. of this set is $2$, which corresponds to Majorana zero modes
  at the domain wall.}}\BibitemShut {Stop}%
\end{thebibliography}

\end{document}